# Fermi Gamma-ray Space Telescope: High-Energy Results from the First Year

P F Michelson[1], W B Atwood[2], S Ritz[2]

1. Department of Physics, Kavli Institute for Particle Astrophysics and Cosmology, and W. W. Hansen Experimental Physics Laboratory, Stanford University, Stanford, CA 94304, USA

2. Santa Cruz Institute for Particle Physics, Department of Physics, University of California at Santa Cruz, Santa Cruz, CA 95064, USA

e-mail: peterm@stanford.edu, atwood@scipp.ucsc.edu, ritz@scipp.ucsc.edu

**Abstract**

The Fermi Gamma-ray Space Telescope (*Fermi*) was launched on June 11, 2008 and began its first year sky survey on August 11, 2008. The Large Area Telescope (LAT), a wide field-of-view pair-conversion telescope covering the energy range from 20 MeV to more than 300 GeV, is the primary instrument on *Fermi*. While this review focuses on results obtained with the LAT, the Gamma-ray Burst Monitor (GBM) complements the LAT in its observations of transient sources and is sensitive to X-rays and γ-rays with energies between 8 keV and 40 MeV. During the first year in orbit, the *Fermi* LAT has observed a large number of sources that include active galaxies, pulsars, compact binaries, globular clusters, supernova remnants, as well as the Sun, the Moon and the Earth. The GBM and LAT together have uncovered surprising characteristics in the high-energy emission of gamma-ray bursts (GRBs) that have been used to set significant new limits on violations of Lorentz invariance. The *Fermi* LAT has also made important new measurements of the Galactic diffuse radiation and has made precise measurements of the spectrum of cosmic-ray electrons and positrons from 20 GeV to 1 TeV.

## 1. Introduction

High-energy γ-ray radiation provides an important astrophysical probe of physical processes in extreme environments and of new physics; *e.g.* particle acceleration to ultra-relativistic energies in the vicinity of black holes, neutron stars, and supernovae remnants and possible signatures of dark matter decay or annihilation. Unlike cosmic rays, once γ-rays emerge from the immediate vicinity of their production they are largely unaffected in the propagation to where they are detected. Naively one might think that γ-rays would traverse inter-galactic space unimpeded, however, if the energy is high enough, these γ-rays can interact with the extragalactic background light (EBL) which pervades the universe. The principal source of this opacity at high energy is pair conversion off the EBL in the infrared-optical-ultraviolet band via $\gamma + \gamma_{EBL} \rightarrow e^+ + e^-$. This process leaves an imprint on the spectrum of distant sources, which, in principle, can tell us about the density of the EBL, and therefore the rate of star formation, versus cosmic time.

In this report we summarize published results from observations made with the Large Area Telescope on the Fermi Gamma-ray Space Telescope (*Fermi*) during the first year of science operations that began in August 2008.[1] High-energy γ-ray observations with *Fermi* were preceded by observations with SAS-II (Fichtel *et al* 1975), COS-B (Bignami *et al* 1975), and EGRET (Fichtel *et al* 1983) on the Compton Gamma-Ray Observatory. The key improvements of the *Fermi* LAT, compared to the previous instruments that have flown, are possible because of the newer technologies, principally for particle tracking and in electronics, that have become available since the construction of EGRET. These key improvements in performance include (i) larger effective area over a much larger field-of-view, (ii) better particle tracking leading to improved angular resolution and background rejection, and (iii) a flexible, fast, multilevel trigger and data acquisition system. The large field-of-view results from the low aspect ratio (height/width) of the LAT made possible by the choice of particle tracking technology (i.e. silicon-strip detectors) that allowed elimination of the time-of-flight triggering system used in EGRET. An overview of the LAT instrument design is given in section 3 of this report.

In the current era, *Fermi* satellite observations up to more than 300 GeV are complemented by ground-based observations, typically above ~100 GeV, with air-Cherenkov telescopes [e.g. H.E.S.S. (Hofmann 1997), VERITAS (Holder *et al* 2006), Magic (Baixeras *et al* 2003), and Cangaroo (Kubo *et al* 2004)] and air-shower arrays such as Milagro (Atkins *et al* 2000). EGRET has provided much of the initial context for *Fermi* observations. For an overview of ground-based, very high-energy γ-ray detectors see Aharonian *et al* (2008a). For a review of EGRET results see Thompson (2008).

The outline of this report is as follows:

1. Introduction

2. Important science objectives addressed by *Fermi*

3. Description of the Fermi Gamma-ray Space Telescope

4. First year of *Fermi* observations

   4.1 Galactic Sources: diffuse radiation, pulsars, globular clusters, supernova remnants, binaries

   4.2 Extragalactic Sources: blazars and active galaxies, radio galaxies, the Large Magellanic Cloud, starburst galaxies, GRBs, diffuse isotropic radiation

   4.3 Local Sources: the Moon, the Sun

   4.4 Dark Matter searches

   4.5 Cosmic-ray electrons and positrons

5. Summary: What next from *Fermi*

---

[1] at https://www-glast.stanford.edu/cgi-bin/pubpub a listing of Fermi LAT Collaboration publications is available

## 2. Important science objectives addressed by *Fermi*

Before launch, it was anticipated that *Fermi* would address a number of important scientific objectives that included the following [The reader is also referred to the more detailed questions listed in section 9 of the EGRET review by Thompson (2008)]:

(1) *Determine the nature of unidentified high-energy γ-ray sources, particularly those seen by EGRET.* The 3$^{rd}$ EGRET Catalog (Hartman *et al* 1999) of 271 sources consists of the single 1991 solar flare bright enough to be detected as a source, the Large Magellanic Cloud (LMC), six pulsars, one probable radio galaxy detection (Cen A), and 66 high-confidence identifications of blazars (BL Lac objects, flat-spectrum radio quasars, or unidentified flat-spectrum radio sources). In addition, 27 lower confidence potential blazar identifications were noted. The catalog also contains 170 sources not firmly identified with known objects. *Fermi* has made significant progress in source identification because of its much narrower point-spread-function (PSF), larger field of view (FoV), and larger effective area, all of which contribute to much smaller error boxes than were possible with EGRET.

(2) *Understand the mechanisms of particle acceleration operating in celestial sources and the origin(s) of cosmic-rays.* EGRET observed high-energy γ-ray emission in several important source categories: active galaxies (AGNs/blazars) containing supermassive black holes ($10^6$-$10^9$ $M_O$), pulsars, the Sun (as well as a small sample of GRBs). There was also reported evidence of emission from supernovae remnants (Sturner and Dermer 1995; Esposito *et al* 1996).

The *Fermi* LAT's wide FoV has allowed AGN/blazar variability to be monitored over a wide range of timescales for a large number of sources (Abdo *et al* 2009a). In these sources, most of the non-thermal γ-ray emission arises from relativistic jets that are narrowly beamed and boosted in the forward direction. There are many questions about the jets, including: how are they collimated and confined? What is the composition both in the initial and the radiative phase? Where does the conversion between the jet's kinetic power and radiation take place? Simultaneous multiwavelength observations of a large number of these sources are crucial for determining the emission mechanisms in order to infer the content of the luminous portions of jets. Also, observations of these extragalactic sources will lead to an understanding of their contribution to the isotropic, apparently diffuse, extragalactic high-energy γ-ray radiation. In the first year of operations, *Fermi* has triggered a number of such observations both in energy bands below that covered by *Fermi* as well as higher-energies accessible with ground-based TeV telescopes. For example, Aharonian *et al* (2009a) have reported on the first simultaneous observations of the blazar PKS 2155-304 that covered optical, X-ray, high-energy (*Fermi* LAT) and very high-energy (e.g. H.E.S.S.) bands.

Pulsars, with their unique temporal signatures, were the only definitively identified EGRET population of Galactic point sources. Aided by their known radio ephemerides, EGRET detected five young radio pulsars at high significance, along with the radio-quiet, isolated X-ray pulsar Geminga and one likely millisecond pulsar (PSR J0218+4232). First year *Fermi* LAT observations have increased the number of known γ-ray pulsars by almost an order of magnitude. The much improved point source sensitivity of the LAT has facilitated successful blind searches for γ-ray pulsars as well as for emission from known radio and X-ray pulsars. Detection of a large number of pulsars by *Fermi*, yielding the pulse profiles and phase-varying spectra of the sources, is providing important insights into understanding these cosmic accelerators.

(3) *Study the high-energy behavior of Gamma-Ray Bursts (GRBs) and transients*. GRBs, that very likely signal the formation of stellar mass black holes, occur at cosmological distances at a rate of a few per day. Before the launch of *Fermi*, it was clear that there are at least two classes of GRBs (Kouveliotou *et al* 1993): long-duration GRBs ($\tau > 2$ s) are associated with low metallicity host galaxies and are usually found in star-forming regions of galaxies while short-duration GRBs can be located in regions of much lower star-formation rate in the host. It is generally thought that long-duration bursts are a result of massive star collapse (Woosley 1993; also see Zhang *et al* 2004 and references therein) and black hole formation while short duration bursts may arise from the coalescence of compact objects (Paczyñski 1986; Bloom *et al* 2006; Nakar 2007).

What was known about high-energy emission from GRBs before *Fermi* came from EGRET observations. There were five GRBs detected with EGRET's spark chamber and calorimeter and additional bursts detected only in the calorimeter because they were outside the spark chamber field-of-view. All of these bursts were accompanied by BATSE (Fishman *et al* 1994) burst detections. Of these bursts, four were clearly long-duration GRBs and one was possibly short-duration. There were two components of high-energy γ-ray emission detected from the long-duration GRBs: >100 MeV emission contemporaneous with the prompt emission detected in the 10 – 1000 keV band, and a delayed component extending to GeV energies that lasted more than an hour in the case of one burst (GRB 940217; Hurley *et al* 1994). Most importantly, EGRET detected one burst (GRB 941017) in which a third power-law component was evident above the commonly used Band function spectrum (Band 1993), with an inferred peak in νF(ν) above 300 MeV during most of the prompt emission phase (Gonzalez *et al* 2003). This indicated that some bursts occur for which the bulk of the energy release falls in the LAT energy band. The spark chamber deadtime (~100 ms/event) of EGRET was comparable to or longer than the pulses in the prompt component of these bursts so EGRET was essentially blind to time structure in the prompt component of the bursts.

*Fermi* has brought a significant new capability for observing the high-energy emission of GRBs. The combination of the Gamma-ray Burst Monitor (GBM) and the LAT allows spectral measurements over seven decades in energy. With a low deadtime (~26 μs/event) the LAT is not severely deadtime limited for the study of the prompt, pulsed component of GRBs. Because very little was known about the nature of GRB high-energy emission, requiring a large extrapolation from previous low-energy measurements (Kaneko *et al* 2006), there was uncertainty in the estimate of the expected number of LAT detections although the estimate of Band *et al* (2009) agrees well with the observed rate of about one per month. The lower-energy GBM rate of 250 detections per year is close to the expected rate.

(4) *Search for signatures of dark matter annihilation or decay and understand the diffuse γ-ray emission from the Galaxy*. EGRET observations were limited in what they could say about this subject, primarily because of the relatively small effective area of the instrument above ~10 GeV due to the self-veto of γ-ray triggers (Thompson *et al* 1993). This effect increased with increasing energy and was caused by backsplash of showering particles generated in the calorimeter that interacted with EGRET's monolithic anticoincidence scintillator shield. The EGRET team did report an apparent excess in the Galactic diffuse emission above around 1 GeV (Hunter *et al* 1997) relative to expectations based on conventional cosmic-ray models but noted that the excess could be due to any of a number of possibilities including the calibration at high energy. The design of the *Fermi* LAT suppresses the self-veto problem and its calibration is supported by extensive accelerator beam tests and a detailed and validated Monte Carlo instrument simulation (Atwood *et al* 2009). Fermi's capabilities for searching for dark matter were widely discussed before launch (e.g., Baltz *et al* 2008; Springel *et al* 2008; Ando and Komatsu 2006; Giocoli, Pieri and Tormen 2008; Kuhlen, Diemand and Madau 2008).

(5) *Determine the attenuation of high-energy γ-rays as a function of cosmological redshift.* Photons with energies above ~10 GeV can probe the era of galaxy formation through absorption by near-UV, optical, and near-IR extragalactic background light (EBL). The EBL in this wavelength band is accumulated radiation from structure and star formation and its subsequent evolution (see, e.g., Madau and Phinney 1996; MacMinn and Primack 1996; Primack *et al* 2001; Hauser and Dwek 2001). Because direct measurements of EBL suffer from large systematic uncertainties due to contamination by the bright foregrounds, blazars and GRBs (with known redshifts) offer an indirect probe via the attenuation caused by absorption of high-energy γ-rays via pair production in the EBL. The *Fermi* LAT energy range extending to greater than 300 GeV offers the opportunity to probe the EBL in the optical-UV band, where absorption breaks are expected for sources located at $z > 0.5$ (Primack *et al* 1999; Salamon and Stecker 1998; Stecker *et al* 2006; Kneiske *et al* 2004; Gilmore *et al* 2009).

## 3. The Fermi Gamma-ray Space Telescope

The Fermi Gamma-ray Space Telescope (*Fermi*), shown in figure 1, has two instruments: (i) the Large Area Telescope (LAT), the primary instrument, is a wide field-of-view imaging telescope covering the energy range ~20 MeV to 300 GeV (Atwood *et al* 2009); (ii) the Gamma-ray Burst Monitor (GBM) is sensitive to X-rays and γ-rays with energies between 8 keV and 40 MeV (Meegan *et al* 2009) and complements the LAT for observations of high-energy transients. Additional information about *Fermi* can be found at the *Fermi* Science Support Center Web site, http://fermi.gsfc.nasa.gov/.

A high-energy γ-ray cannot be reflected or refracted and instead interacts with matter primarily by conversion into an $e^+e^-$ pair. The LAT therefore is a pair-conversion telescope with a precision converter-tracker section followed by a calorimeter. These two subsystems each consist of a 4 x 4 array of 16 modules (see figure 1). The active elements of the tracker are silicon-strip detectors. The calorimeter is a hodoscopic configuration of 8.6 radiation lengths of CsI crystals that allows imaging of the shower development in the calorimeter and thereby corrections of the energy estimate for the shower leakage fluctuations out of the calorimeter. The total thickness of the tracker and calorimeter is approximately 10 radiation lengths at normal incidence. A segmented anticoincidence detector (ACD) covers the tracker array, and a programmable trigger and data acquisition system uses prompt signals available from the tracker, calorimeter, and ACD to form a trigger that initiates readout of these three subsystems. The onboard trigger is optimized for rejecting events triggered by cosmic-ray background particles while maximizing the number of events triggered by γ-rays, which are transmitted to the ground for further processing.

The detailed performance specifications of the LAT can be found in Atwood *et al* (2009) while the specifications of the GBM are in Meegan *et al* (2009). Table 1 summarizes the performance characteristics of the LAT.

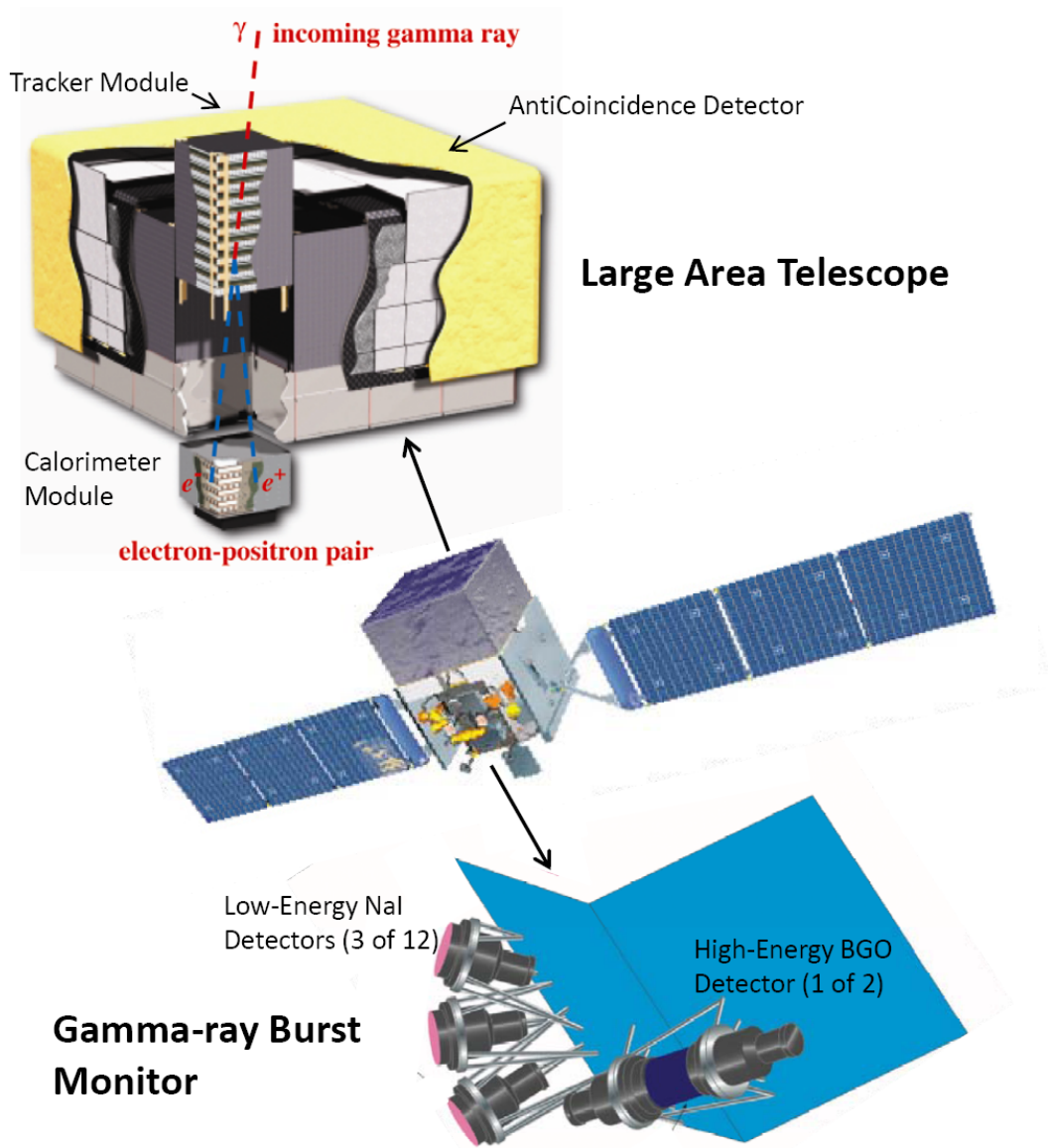

**Figure 1.** The Fermi Gamma-ray Space Telescope and its two instruments. The Large Area Telescope (LAT) images the sky in the energy band from ~20 MeV to more than 300 GeV while the Gamma-ray Burst Monitor (GBM) complements the LAT for the study of GRBs and transients, providing spectral coverage from 8 keV to about 40 MeV.

The GBM consists of two sets of six low-energy (8 keV to 1 MeV) NaI(Tl) detectors and a high-energy (0.2 to 40 MeV) BGO detector. These sets of detectors are mounted on opposite sides of the spacecraft as indicated in figure 1. The GBM detectors are unshielded and uncollimated scintillation detectors distributed around the spacecraft with different viewing angles in order to determine the direction of a burst by comparing the count rates of different detectors. The GBM communicates burst positions determined on-board to the LAT. For bursts above a preset threshold, the spacecraft is autonomously re-pointed to keep the GRB within the LAT FoV for the next 5 hours allowing optimal observation of temporally extended high-energy γ-ray emission with the LAT.

**Table 1.** Performance characteristics of the *Fermi* LAT[2].

| Parameter | Value or Range |
|---|---|
| Energy range | 20 MeV – 300 GeV |
| Effective area at normal incidence | ≤8,400 cm$^2$ |
| Energy resolution (eq. Gaussian 1σ; on-axis): | |
| 100 MeV – 1 GeV | 15% - 9% |
| 1 GeV – 10 GeV | 8% - 9% |
| 10 GeV – 300 GeV | 8.5% - 18% |
| Single photon angular resolution (space angle): | |
| > 10 GeV | <0.15$^\circ$ |
| 1 GeV | 0.6$^\circ$ |
| 100 MeV | 3.5$^\circ$ |
| Field of View (FoV) | 2.4 sr |
| Timing accuracy | 300 ns |
| Single event readout time (dead time) | 26.5 µs |

To take full advantage of the LAT's large FoV, the primary observing mode of *Fermi* is the so-called "scanning" mode (sometimes called "rocking" mode) in which the normal to the front of the instrument is offset (typically 0° to 60°) from the zenith and towards the pole of the orbit on alternate orbits and in the opposite direction from the zenith for the other orbits. Thus, after two orbits (or about 3 hours for *Fermi*'s orbit at 565 km and 25.6° inclination) the sky exposure is almost uniform. For autonomous repoints or for other targets of opportunity, the observatory can be inertially pointed.

---

[2] up-to-date LAT performance specifications can be found online at
http://www-glast.slac.stanford.edu/software/IS/glast_lat_performance.htm

## 4. First year of Fermi LAT observations

During the first year of operation, *Fermi* detected more than 150 million γ-rays, compared with EGRET that detected about 1.4 million γ-rays during nine years of operation. Figure 2 shows the summed map of γ-rays detected by *Fermi* above 200 MeV, in Galactic coordinates. The bright band of emission running horizontally across the center of the figure is primarily diffuse emission from the disk of the Milky Way. The Galactic center is at the center of the map.

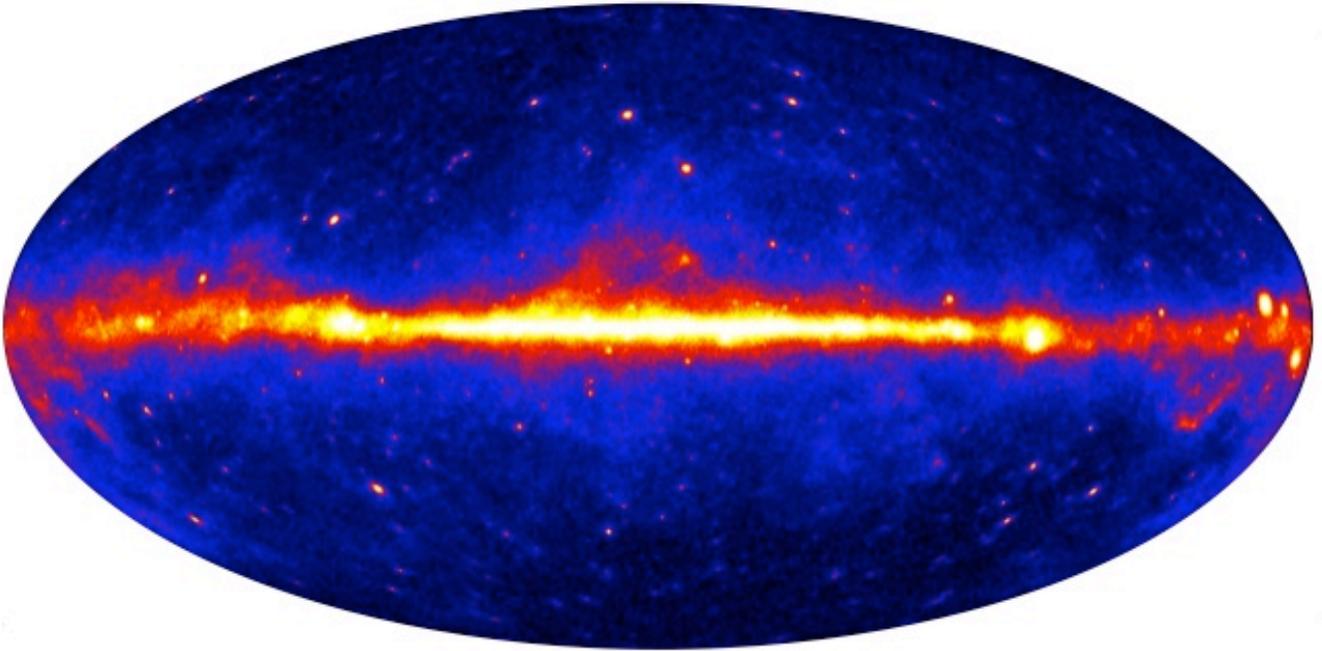

**Figure 2.** False color image of the γ–ray sky above 200 MeV as seen by *Fermi* from one year of observation. The map is in Galactic coordinates with the center of the Galaxy at the center of the map. The bright band of emission from the Milky Way is dominated by diffuse emission along the plane of the Galaxy.

The data used to produce this image contains a wealth of new information about the γ-ray sky that is discussed in the following sections. The *Fermi* LAT Collaboration has already published a list of 205 bright high-energy γ-ray sources (significance > 10 σ), the Bright Source List (BSL), based on the first 3 months of the first year all-sky survey (Abdo *et al* 2009b). This list will soon be followed by the first *Fermi* LAT source catalog based on the first 11 months of the all-sky-survey. Of the 205 BSL sources, 13 were found to have potential associations with sources near pulsar wind nebulae (PWNs) or near supernova remnants (SNRs) while 37 had no associations. The remaining 155 BSL sources have firm identifications as follows: 30 γ-ray pulsars, 2 high-mass X-ray binaries, 46 BL Lac blazars, 64 flat-spectrum radio quasars (FSRQs), 9 other blazars, 2 radio galaxies, 1 globular cluster, and the Large Magellanic Cloud (LMC).

*4.1 Galactic sources*

*4.1.1 Galactic Diffuse Emission:* High-energy γ-ray emission from the Milky Way is dominated by diffuse emission that is particularly bright along the plane of the Galaxy and most pronounced toward the inner part of the Galaxy. The Galactic diffuse emission is generated primarily by energetic cosmic rays (*e.g.* electrons and protons) that interact with interstellar gas (via $\pi^o$ production and bremsstrahlung) and interstellar radiation fields (via inverse Compton scattering; Stecker 1977). EGRET measurements of the diffuse radiation indicated excess γ-ray emission above ~1 GeV relative to conventional Galactic diffuse emission models that are consistent with the locally measured cosmic-ray spectra; the so-called "EGRET GeV excess". De Boer (2005) and others proposed that the excess, observed in all directions on the sky, was due to γ-rays from annihilating dark matter. Before the launch of *Fermi*, this explanation was challenged on a number of grounds including that the model was inconsistent with the measured flux of antiprotons (Bergström *et al* 2006). A more mundane possibility is that the excess reflects a miscalibration of EGRET at high energies (Stecker *et al* 2008; Thompson 2008).

Based on analysis of data from the first year of observations, the *Fermi* LAT collaboration reported on measurements of the diffuse γ-ray emission from 100 MeV to 10 GeV and Galactic latitudes $10^o \leq |b| \leq 20^o$ (Abdo *et al* 2009c). A comparison of the EGRET and *Fermi* LAT diffuse spectra is shown in the left panel of figure 3. The LAT spectrum is softer above 1 GeV and is consistent with a model for diffuse emission that reproduces the local cosmic-ray spectrum and, at least over this part of the energy spectrum, does not require an additional component.

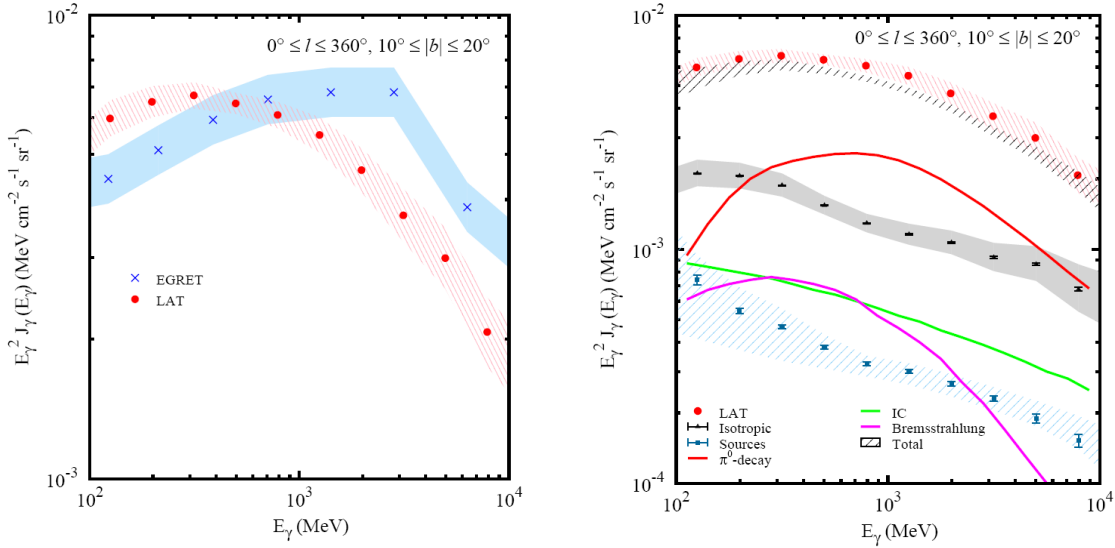

**Figure 3.** Left: Diffuse emission intensity averaged over all Galactic longitudes for latitude range $10^o \leq |b| \leq 20^o$. Red dots, *Fermi* LAT; blue dots, EGRET. Systematic uncertainties: red, Fermi LAT; blue, EGRET. Right: LAT data with diffuse model, point sources, and Unidentified background (UIB) components. The UIB, consisting of an extragalactic diffuse component, emission from unresolved sources, and residual particle backgrounds, was determined by fitting the data and sources over all Galactic longitudes $|b| > 30^o$ for the full energy range shown. (Reprinted by permission from the American Physical Society: *Physical Review Letters* (Abdo *et al* 2009c), copyright 2009.)

*4.1.2 Pulsars:* The observations of γ-ray emission from pulsars began with SAS-2 and COS-B observations of pulsed γ-ray emission from the Crab and Vela. EGRET increased this number to 6. Thompson (2004, 2008) provides a summary of the EGRET observations of pulsars.

Extensive radio and X-ray timing observations of known pulsars (Smith *et al* 2008) and the development of an efficient time-differencing algorithm for periodicity searches in very sparse γ-ray data (Atwood *et al* 2006), along with the much improved performance of the LAT, has led to the discovery of an order of magnitude more γ-ray pulsars. The total number of high-energy γ-ray pulsars detected with high confidence stands at 46. These include the 6 EGRET pulsars, 16 discovered in blind searches (Abdo *et al* 2009d) and another 24 that are also radio pulsars whose pulsed γ-ray emission was discovered by using the known radio timing ephemerides. Of the radio pulsars, 8 are millisecond pulsars (Abdo *et al* 2009e). 22 of the 46 *Fermi* pulsars were EGRET sources, 16 of which were unidentified sources in the 3$^{rd}$ EGRET catalog.

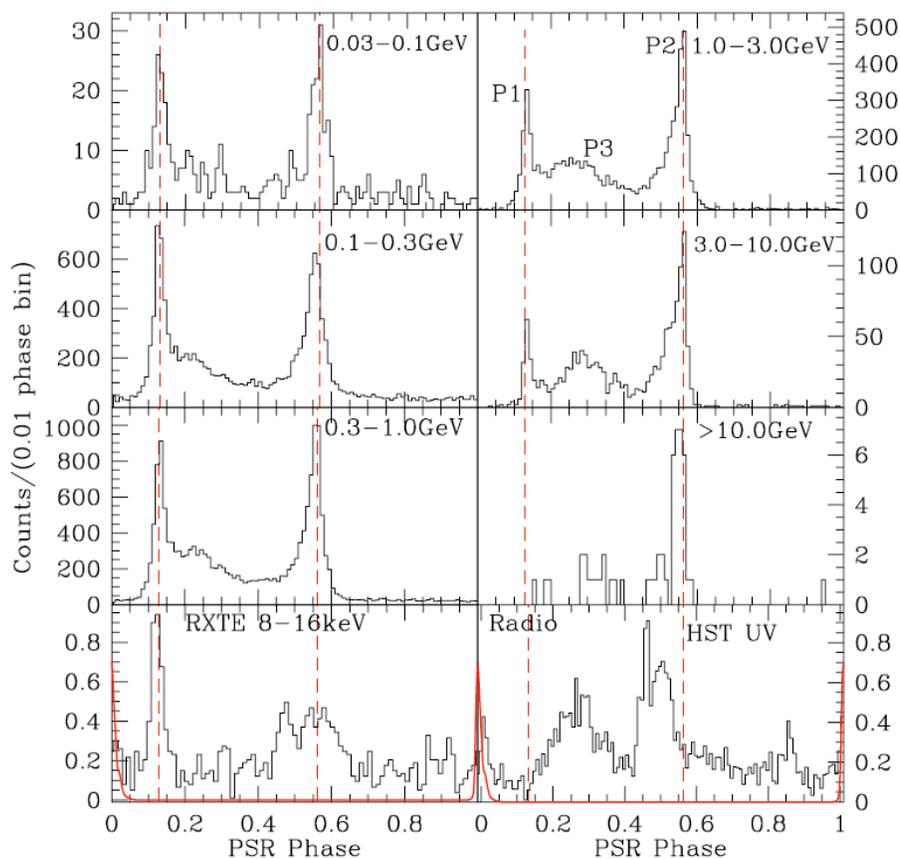

**Figure 4.** Evolution of the Vela γ-ray pulse profile over 3 decades of energy. Dashed lines show the phases of the P1 and P2 peaks determined from the broadband γ-ray light curve. The main peaks P1, P2, and P3 are labeled in the top light panel. The bottom left panel shows the 8-16 keV *RXTE* pulse profile of Harding *et al* (2002) along with the radio pulse profile (in red). The 4.1-6.5 eV *HST*/STIS NUV pulse profile of Romani *et al* (2005) is shown in the lower right panel. (Reprinted by permission from the American Astronomical Society: *Astrophysical Journal* (Abdo *et al* 2009f), copyright 2009.)

γ-rays are produced by particles that are accelerated to high energies in the magnetospheres of neutron stars through a combination of curvature radiation, synchrotron radiation and inverse Compton scattering. The observed energy spectra depend on the rotational phase and contain information about the physical mechanisms producing the radiation and the location of the emission. Figure 4 shows the evolution of the Vela γ-ray light curve in 6 energy bins covering 3 decades in energy, determined from LAT observations during the first 2.5 months in orbit (Abdo *et al* 2009f). The phase-averaged γ-ray energy spectrum measured by the LAT is well represented by a power law with an exponential cutoff and excludes a hyper-exponential cutoff as would be expected for models radiating the γ-rays from the surface near the magnetic polar cap of the neutron star, thus favoring outer-magnetosphere emission models.

The first blind search discovered γ-ray pulsar, LAT PSR J0007+7309 found in the supernovae remnant CTA1 (Abdo *et al* 2008), is illustrative. Figure 5 shows the γ-ray pulse profile and the 95% error circle of the LAT pulsar superposed on a 1420 MHz radio map (Pineault *et al* 1997) of the shell supernova remnant CTA1 along with the much larger error circle of the corresponding EGRET source 3EG J0010+7309. The γ-ray pulsar is coincident with an X-ray point source RX J00070+7302.

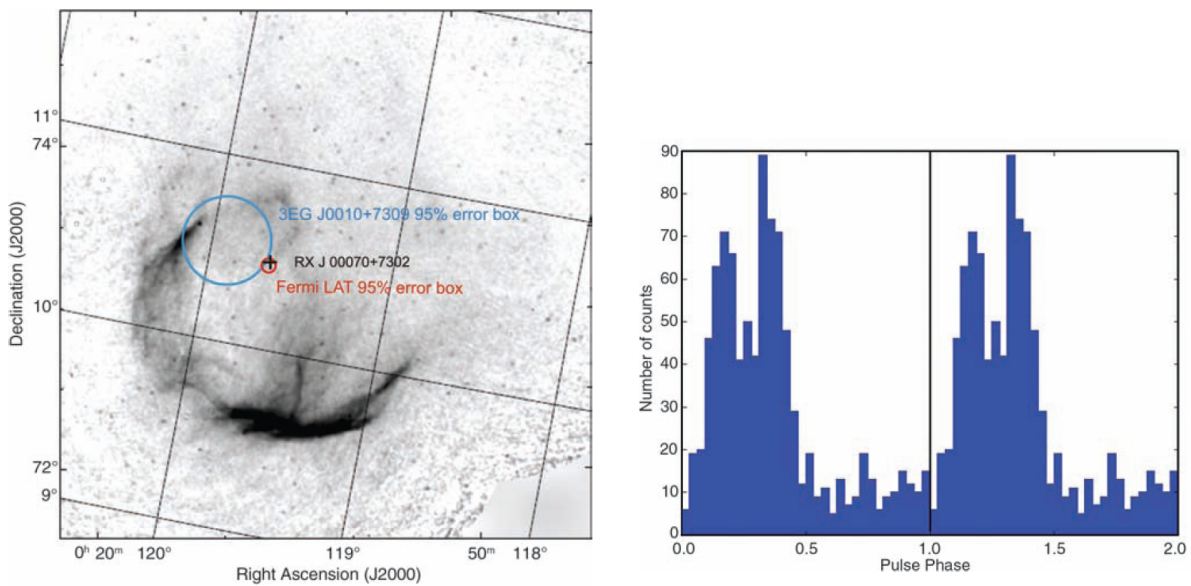

**Figure 5.** Left: The Fermi LAT discovery of the pulsar PSR J0007+7309 in the supernovae remnant CTA1 in data from the first few weeks of science operations. The LAT 95% error region and that of the corresponding EGRET source 3EG J0010+7309 are superposed on a 1420 MHz radio map. Right: The γ-ray light curve (> 100 MeV) of the pulsar shown over two rotation periods using data from the first 2.5 months of Fermi observations. The pulsar has a period of 316.86 milliseconds and a period derivative of $3.614 \times 10^{-13}$ seconds per second. Two maxima in the broad emission feature are separated by ~ 0.2 in phase. (Reprinted by permission from the American Association for the Advancement of Science: *Science* (Abdo *et al* 2008), copyright 2008.)

The measured period and period derivative of this pulsar give an estimate of the total energy loss rate of the pulsar, the so-called spin-down power, of ~4.5 x $10^{35}$ ergs s$^{-1}$, which is sufficient to power the synchrotron pulsar wind nebula (PWN) embedded in the supernova remnant (SNR) (Abdo *et al* 2008). Also, the inferred age of the pulsar, 14,000 years, is consistent with the estimated age of the SNR.

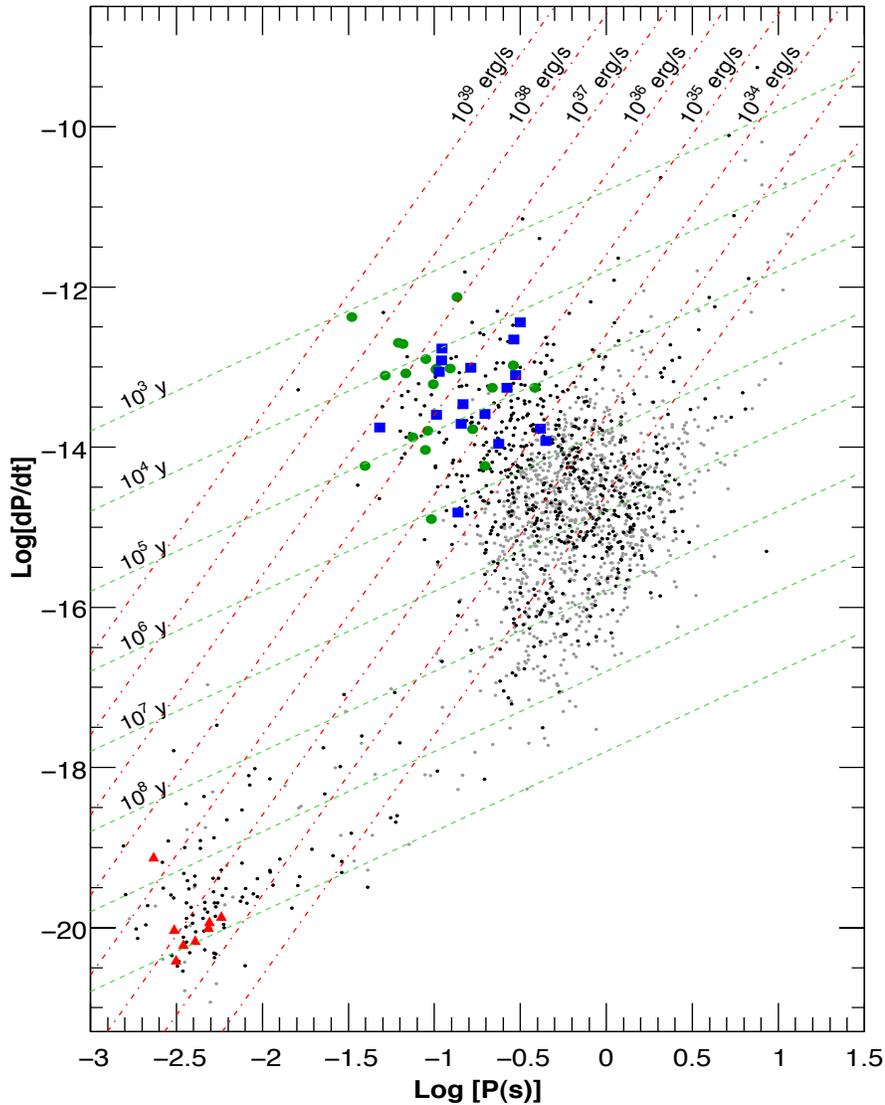

**Figure 6.** Period – period derivative distribution of the γ-ray pulsars detected by the *Fermi* LAT during the first year sky survey (Abdo *et al* 2009g). Triangles – millisecond γ-ray pulsars; squares – γ-ray pulsars detected with blind search; circles – non-millisecond γ-ray pulsars detected using known radio ephemerides. The dots are the ~1800 pulsars in the ATNF catalog (Manchester *et al* 2005).

Figure 6 shows the periods and period derivatives of all of the 46 *Fermi* high-confidence γ-ray pulsars. The characteristic age and the spin-down luminosity of each pulsar can be read from the figure. Except for the much older population of millisecond pulsars, the characteristic ages of the pulsars lie in the range $10^3$ to $10^6$ years and the spin-down luminosities are in the range ~3 x $10^{33}$ to 4.6 x $10^{38}$ ergs s$^{-1}$. A power law with an exponential cutoff can describe nearly all of the energy spectra of these pulsars, with cutoff energies in the range from just under 1 GeV to several GeV, suggesting that for most of the pulsars the γ-ray emission comes mainly from the outer magnetosphere (Abdo *et al* 2009g). Roughly 75% of the γ-ray pulse profiles show two peaks, whereas the radio pulse profiles of more than 70% of young pulsars have one peak (Abdo *et al* 2009g).

*4.1.3 Globular Clusters:* Globular clusters are among the oldest constituents of our Galaxy. They contain a relatively large fraction of close binary systems, many low-mass X-ray binary systems with neutron stars, and they contain many millisecond pulsars (MSPs). Before *Fermi* they had been detected in all bands of the electromagnetic spectrum except for γ-rays.

Of the 8 millisecond γ-ray pulsars detected by *Fermi*, most are just a few hundred parsecs from the sun, implying their γ-ray luminosities generally do not exceed $10^{33}$ erg s$^{-1}$ (Abdo *et al* 2009e). Since the nearest globular cluster is several kiloparsecs away, it is unlikely, but not impossible, that *Fermi* will detect individual MSPs in globular clusters. However, individual globular clusters contain tens to hundreds of MSPs so the possibility of detecting the cumulative emission from MSPs in globular clusters is likely.

47 Tucanae (NGC 104) was one of the most promising candidates for detection by *Fermi* because of the 23 known MSPs and its relative proximity (4kpc) (McLaughlin *et al* 2006). Indeed, it has been detected from *Fermi* LAT observations made during the first year sky survey (Abdo *et al* 2009h). Figure 7 shows a γ-ray image of a region centered on the position of 47 Tuc and the spectral energy distribution of the source. The 95% confidence region for the location of the γ-ray source coincides with the core region of 47 Tuc. The spectrum from 200 MeV to 20 GeV is well fit by a power law with an exponential cutoff, consistent with what might be expected from a collection of pulsars. Assuming the γ-ray efficiencies of the MSPs in 47 Tuc are equal to those of the nearby galactic field sample and that their average geometrical beaming correction factors are the same, Abdo *et al* (2009h) have estimated that the number of γ-ray pulsars in 47 Tuc lies between 7 and 62 (95% confidence interval). In any case, it seems very likely that MSPs are the primary population of γ-ray sources in 47 Tuc and in globular clusters in general.

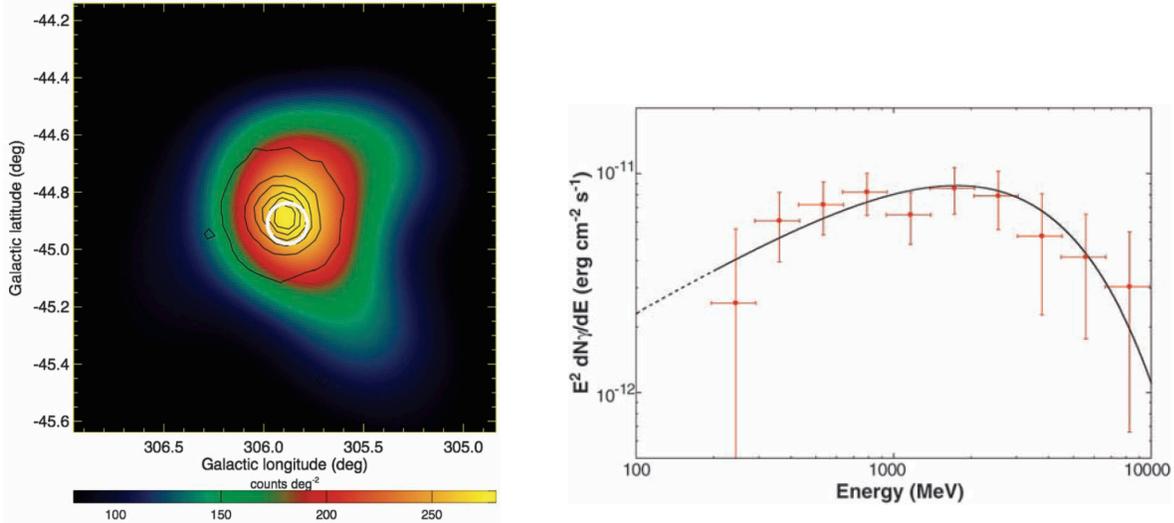

**Figure 7.** Left: *Fermi* LAT γ-ray image (200 MeV to 10 GeV) of a 1.5° x 1.5° region centered on the position of 47 Tuc. A total of 290 counts were detected from the γ-ray source. The map has been adaptively smoothed, imposing a minimum signal-to-noise of 5. Black contours indicate the stellar density in 47 Tuc as derived by McLean *et al* (2000). The white circle shows the 95% confidence region for the location of the *Fermi* source. The γ-ray source coincides with the core region of 47 Tuc. Right: Spectral energy distribution of the *Fermi* source in 47 Tuc. The solid line shows the fit of an exponentially cut-off power law from 200 MeV to 20 GeV. (Reprinted by permission from the American Association for the Advancement of Science: *Science* (Abdo *et al* 2009h), copyright 2009.)

*4.1.4 Supernova Remnants:* The *Fermi* LAT Collaboration has reported the detections of high-energy γ-ray emission from several supernova remnants (SNRs) that include both the middle-aged remnants ($\sim 10^4$ yr) W51C, W44, W28 and IC443 and the young remnants ($< \sim 10^3$ yr) Cassiopeia A and RXJ 1713.7-3946 (Funk 2009). In addition, Slane (2009) has recently reported the high-energy detection of the middle-aged remnant G349.7+0.2. For RXJ 1713.7–3946, W51C, W44, and IC443, the spatial extent of the GeV emission has been resolved. Previous observations with EGRET found some γ-ray sources near radio-bright SNRs (Esposito *et al* 1996) but the possible origins of the EGRET sources were not clear, mainly due to relatively poorer localizations.

For some time it has been thought that supernovae produce cosmic rays (e.g. Hayakawa 1956) with galactic cosmic ray particles produced by acceleration in the expanding shocks of SNRs through diffusive shock acceleration (e.g., Blandford and Eichler 1987). *Fermi* observations, together with recent observations of synchrotron X-rays and very high-energy γ-rays (e.g., Reynolds 2008; Aharonian *et al* 2008a) in young SNRs, have strengthened this conjecture. A key requirement is that the efficiency for converting the kinetic energy of supernovae explosions into the energy of relativistic electrons and nuclei be relatively high, of order 10% (Ginzburg and Syrovatskii 1964), in order to explain the observed flux of galactic cosmic rays.

The middle-aged remnants detected by *Fermi* are in close proximity to molecular clouds. These sources provide the opportunity to identify the $\pi^0$-decay $\gamma$-rays that should be produced when the nuclear cosmic rays generated in the SNR interact with a nearby molecular cloud (Stecker 1976).

W51C, the first reported SNR detection with *Fermi*, is an interesting case (Abdo *et al* 2009i). This source is a radio-bright SNR, located a distance of about 6 kpc away, that has an extended elliptical shape of 50' x 38'. The spatial distribution of $\gamma$-rays detected with Fermi is significantly extended on the scale of $0.22^O$. The luminosity in the band 0.2-50 GeV is $10^{36}$ erg s$^{-1}$. The source has also been detected in the TeV band by H.E.S.S. (Fiasson *et al* 2009). The Fermi LAT Collaboration has reported that the LAT data suggest that $\pi^0$-decay is the dominant contribution to the $\gamma$-ray signal from this source (Abdo *et al* 2009i) suggesting in turn that accelerated ions are indeed produced in W51C.

Recently, Hewitt *et al* (2009) have pointed out that OH 1720 MHz maser emission from SNRs correlates with $\gamma$-ray emission. Indeed, all of the middle-aged remnants detected by *Fermi* also exhibit maser emission. OH 1720 MHz maser emission in these systems is believed to result from shock waves propagating perpendicular to the line of sight, hitting nearby molecular clouds that have sufficient density (Lockett *et al* 1999). This tends to reinforce the picture that in these remnants the $\gamma$-ray emission is from the decay of $\pi^o$s produced by the interaction of high-energy protons accelerated in the SNR with a nearby dense molecular cloud.

*4.1.5 Binary Sources*: Beginning with COS-B and EGRET observations, the high-mass X-ray binaries (HMXBs) LS I +61$^o$303 and LS 5039 have been thought to be associated with high-energy $\gamma$-ray sources; e.g. the COS-B source 2CG 135+01 (Hermsen *et al* 1977) and the EGRET source 3EG J0241+6103 (Kniffen *et al* 1997) in the case of LS I +61$^o$303 and 3EG J1824+1514 in the case of LS 5039 (Parades *et al* 2000). Variability in the EGRET light curve of LS I +61$^o$303 could not be firmly established (Tavani *et al* 1998; Nolan *et al* 2003), particularly on the timescale of the orbital period of 26.4960 ± 0.0028 days. TeV emission has been detected from both of these sources, showing orbitally modulated emission for LS I +61$^o$303 (Albert *et al* 2006; Acciari *et al* 2008) and periodic emission for LS 5039 (Aharonian *et al* 2006). TeV emission has also been observed from the binary system powered by the radio pulsar PSR B1259-63 (Aharonian *et al* 2005) and from Cygnus X-1 (Albert *et al* 2006), a system that likely contains a black hole.

*Fermi* LAT observations from 2008 August to 2009 March indicate that the high-energy emission (20 MeV – 100 GeV) from LS I +61$^o$303 is orbitally modulated at 26.6 ±0.5 days (Abdo *et al* 2009j). Figure 8 shows the power spectrum of the $\gamma$-ray light curve and the folded light curve binned in orbital phase (Abdo *et al* 2009j).

Abdo *et al* (2009j) note that the folded *Fermi* light curve peaks around phase 0.3, just after periastron passage (when the compact object is closest to the Be star companion), in contrast to the behavior above 100 GeV where the peak emission occurs at phases 0.6-0.7, near apastron.

The high-energy emission from LS I +61$^o$303 likely arises from inverse Compton scattering from stellar photons of a population of electrons accelerated in the vicinity of the compact object, be it a neutron star or a black hole. In this case, the fact that the *Fermi* flux peaks close to periastron is consistent with inverse Compton scattering from electrons close to the compact object (Abdo *et al* 2009j). An independently varying spectral component is needed to explain why the TeV emission peaks at a different orbital phase.

Abdo *et al* (2009k) have also reported variability from LS 5039 that is consistent with the binary period, with the emission being modulated at 3.903±0.005 days. The light curve exhibits a broad peak around superior conjunction that is in agreement with inverse Compton scattering models. The spectrum of the source is well represented by a power law with a cutoff at about 2.1 GeV, suggestive of magnetospheric emission similar to that seen in many of the pulsars observed by *Fermi*.

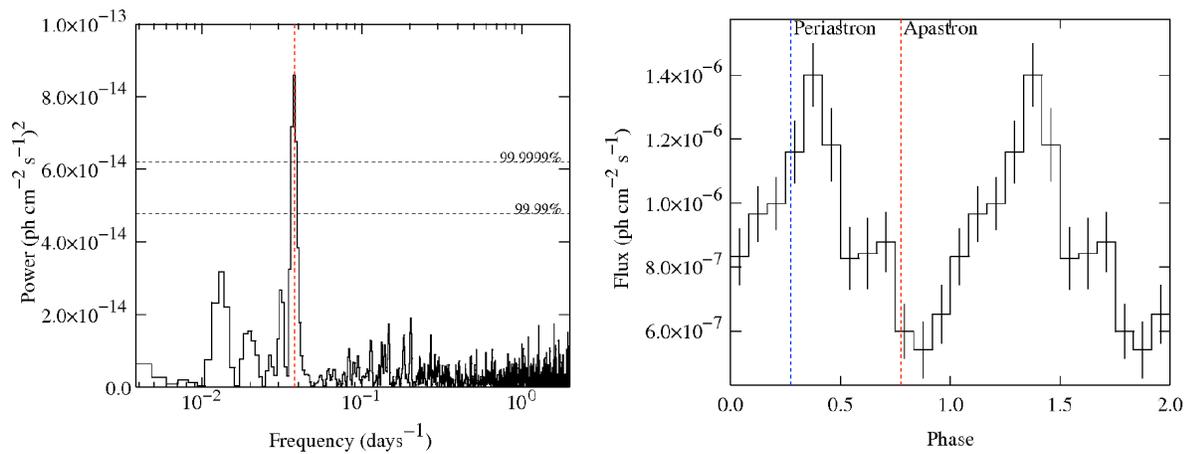

**Figure 8.** Left: Power spectrum of LS I +61°303 γ-ray light curve. The vertical dashed line indicates the known orbital period from Gregory (2002) that coincides with the peak in the power spectrum. The horizontal dashed lines indicate significance levels. Right: Folded light curve binned in orbital phase. Dashed lines indicate periastron and apastron as given by Aragona *et al* (2009). (Reprinted by permission from the American Astronomical Society: *Astrophysical Journal* (Abdo *et al* 2009j), copyright 2009.)

The Fermi LAT Collaboration has also reported LAT observations of modulated high-energy γ-ray emission from the microquasar Cygnus X-3 (Abdo *et al* 2009l). The γ-ray source exhibits variability that is correlated with radio emission from the relativistic jets of Cygnus X-3. The identification of this source is made secure by the detection of modulation of the γ-ray emission at the 4.8 hour orbital period of Cygnus X-3. Tavani *et al* (2009) have also reported *Agile* detection of γ-ray flares from this source that are consistent with those observed with Fermi.

*4.2 Extragalactic sources:*

*4.2.1 Blazars and Active Galaxies:* Blazars are a class of active galaxies in which the detected emission is dominated by non-thermal radiation generated in a relativistic jet flowing out from the region near a central black hole. The angle between the line-of-sight and the axis of the jet is typically a few degrees or less in these sources resulting in the superluminal motion often observed in VLBI radio observations of blazars (e.g., Hughes 1991). The detection of short-term variability of γ-ray emission from the blazar 3C279 by EGRET on a scale of days (Kniffen *et al* (1993) reinforced the idea that jets were also a source of high-energy emission. The argument made supporting this conclusion is roughly as follows: rapid variability requires a compact emitting region, and such a region should be opaque to γ-rays due to γ–γ pair production, the interaction of a high-energy γ-ray with a lower-energy photon to produce an electron–positron pair, $\gamma + \gamma = e^{+} + e^{-}$, unless most of the photons are moving in the same direction, as in a jet. The jet model for the emission was also supported by EGRET observations of correlated variability of γ-ray flares with flares seen at other wavelengths (e.g., PKS 1406−076, Wagner *et al* 1995). Such correlated variability is a valuable source of information about how such jets are formed and their composition, how they are collimated and how they carry energy.

Most blazars are further observationally classified as Flat Spectrum Radio Quasars (FSRQs) or BL Lacertae (BL Lac) objects. BL Lacs in turn are often classified as low-energy-peaked BL Lac objects (LBLs) and high-energy-peaked BL Lac objects (HBLs). While the term 'blazar' is not uniquely defined, it typically includes active galactic nuclei that are radio-loud, with flat radio spectrum, and exhibit polarization in optical and/or radio as well as significant variability. Blazars typically have a spectral energy distribution (SED) with two very broad peaks. At lower frequencies, from radio to optical or sometimes X-rays, the emission is thought to be dominated by synchrotron radiation of high-energy electrons in the jet. The higher energy peak, extending from X-rays upward, is thought to result from inverse Compton scattering of low-energy photons by the same population of high-energy electrons that produces the lower-energy synchrotron radiation. The source of the photons to be upscattered can be the synchrotron radiation itself (synchrotron self-Compton) or some outside source of photons (external Compton).

The first three months of sky-survey operation with the LAT revealed 132 bright sources at $|b| > 10^{o}$ with significance greater than 10σ (Abdo *et al* 2009a). High-confidence associations with known active galactic nuclei (AGNs) have been made for 106 of these sources. This sample is referred to as the LAT Bright AGN Sample (LBAS). It contains two radio galaxies, namely, Centaurus A and NGC 1275, and 104 blazars consisting of 58 FSRQs, 42 BL Lac objects, and 4 blazars with unknown classification. Four new blazars were discovered on the basis of the LAT detections. The LBAS includes 10 HBLs, sources which were previously difficult to detect in the GeV range. This is primarily due to the large increase in effective area and the narrower PSF of the LAT relative to EGRET at high energies. Another 10 lower-confidence associations were also found. Only 33 of the sources, plus two at $|b| < 10^{o}$, were previously detected with EGRET, probably because of variability. By comparison, the Third EGRET catalog of high-energy γ-ray sources contains 66 high-confidence blazars, of which ~77% are identified as FSRQs and ~23% are identified as BL Lac objects, compared with nearly 40% of the LBAS sources being BL Lacs. Figure 9 shows the redshift distributions for LBAS FSRQs and BL Lac objects, compared with the parent population distributions in the BZCat catalog (Massaro *et al* 2009).

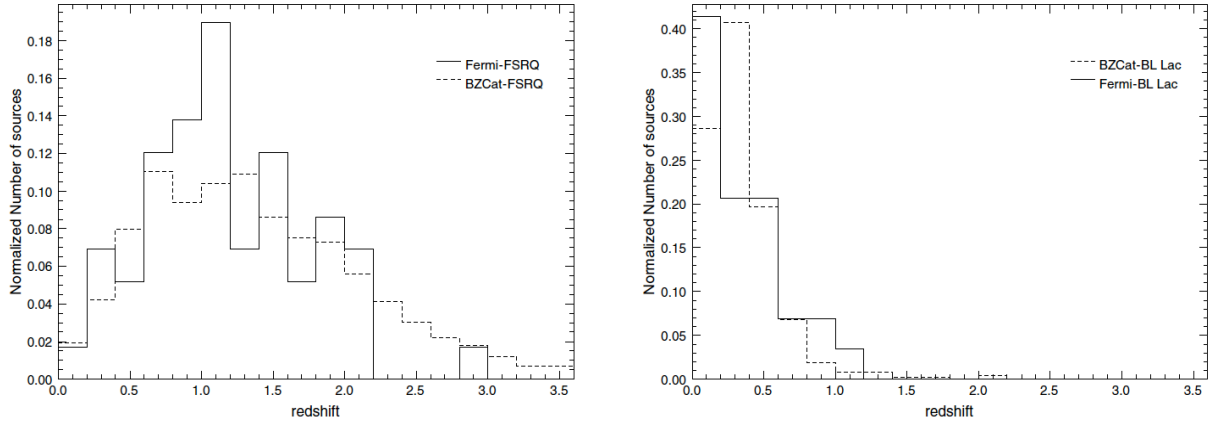

**Figure 9.** Left: Redshift distribution for the FSRQs in the LBAS (solid) and in the BZCat catalog (dashed). Right: Redshift distribution for the BL Lacs in the LBAS (solid) and in the BZCat catalog (dashed). (Reprinted by permission from the American Astronomical Society: *Astrophysical Journal* (Abdo *et al* 2009a), copyright 2009.)

The analysis of the γ-ray properties of the LBAS sources reveals that the average GeV spectra of BL Lac objects are significantly harder than the spectra of FSRQs. The top panel of figure 10 shows the photon power-law index distribution for all LBAS sources. This distribution is similar to that observed for the EGRET sample (Nandikotkur *et al* 2007): it is roughly symmetric and centered at $\gamma = 2.25$. The corresponding distributions for FSRQs and BL Lacs are shown in the bottom and middle panels of figure 10, respectively. These distributions are clearly distinct, with little overlap between them (Abdo *et al* 2009a). Although indications of the existence of two spectrally distinct populations (BL Lacs and FSRQs) in the EGRET blazar sample were mentioned in the literature (Pohl *et al* 1997; Venters and Pavlidou 2007), the *Fermi* LAT observations are the first time that the distinction appears so clearly.

A marginal correlation between radio and peak γ-ray fluxes is observed. Unlike surveys at optical or X-ray energies in which the majority of AGNs are radio-quiet (e.g. della Ceca *et al* 1994; Ivezic *et al* 2002), all the Fermi LAT AGNs, like the EGRET AGNs, are strong radio sources, and most exhibit superluminal motion (Jorstad *et al* 2001; Kellermann *et al* 2004; Lister *et al* 2009).

Abdo *et al* (2009a) also constructed luminosity functions to investigate the evolution of the different blazar classes, with positive evolution indicated for FSRQs but none for BL Lacs.

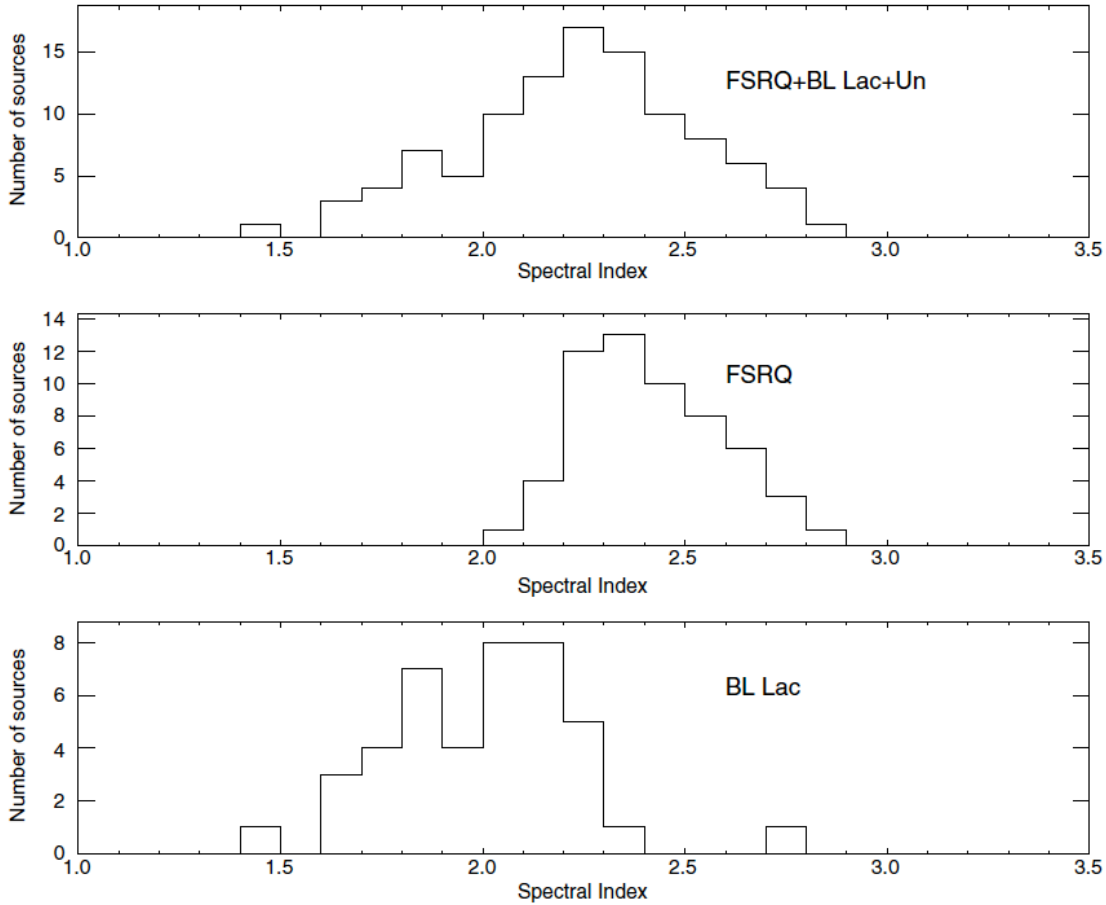

**Figure 10.** Photon spectral index distributions for the bright *Fermi* LAT blazars. Top: all sources. Middle: FSRQs. Bottom: BL Lacs. (Reprinted by permission from the American Astronomical Society: *Astrophysical Journal* (Abdo *et al* 2009a), copyright 2009.)

While a simple power-law provides a good description of the SED for many blazar sources over the energy range covered by the LAT, this is not always the case. In particular, *Fermi* LAT observations of 3C 454.3, figure 11, covering 2008 July 7-October 6, indicated strong, highly variable γ-ray emission (Abdo *et al* 2009m). The observed γ-ray spectrum, figure 12, is not consistent with a simple power law, but instead steepens strongly above ~2 GeV, and is well described by a broken power law with photon indices of ~2.3 and ~3.5 below and above the break, respectively. This is the first direct observation of a break in the spectrum of a high-luminosity blazar above 100 MeV, and is interpreted by Abdo *et al* (2009m) as either likely due to an intrinsic break in the energy distribution of the radiating particles or, possibly, the spectral softening above 2 GeV could be due to γ-ray absorption via photon–photon pair production on the soft X-ray photon field of the host active galactic nucleus, but such an interpretation would require the dissipation region to be located very close (~100 gravitational radii) to the black hole, which would be inconsistent with the X-ray spectrum of the source.

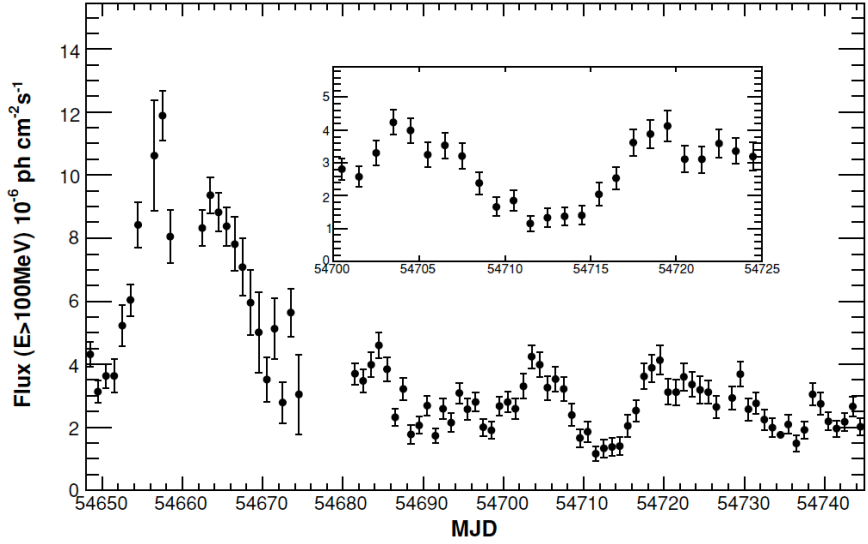

**Figure 11.** Flux light curve of 3C 454.3 in the 100MeV–300 GeV band. The LAT operated in survey mode throughout these observations except during the period MJD 54654–54681 (2008 July 7–August 2), when it operated in pointed mode. The inset shows a blow up of the period MJD 54700–54725. The error bars are statistical only. (Reprinted by permission from the American Astronomical Society: *Astrophysical Journal* (Abdo *et al* 2009m), copyright 2009.)

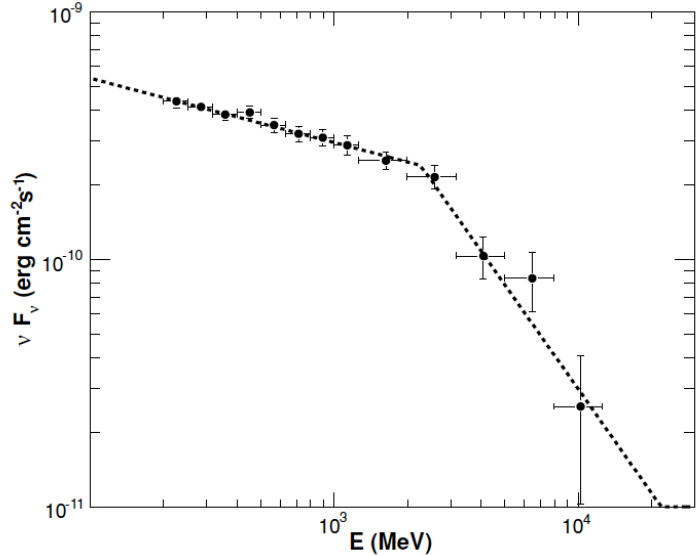

**Figure 12.** $vF_v$ distribution of the summed *Fermi* LAT observations of 3C 454.3 for the 2008 August 3–September 2 time span. The model, fitted over the 200 MeV–300 GeV range, is a broken power law with photon indices $\Gamma_{low} = 2.27\pm0.03$, $\Gamma_{high} = 3.5\pm0.3$, and a break energy $E_{br} = 2.4\pm0.3$ GeV, and the apparent isotropic $E > 100$ MeV luminosity of $4.6 \times 10^{48}$ erg cm$^{-2}$ s$^{-1}$. The error bars are statistical only. (Reprinted by permission from the American Astronomical Society: *Astrophysical Journal* (Abdo *et al* 2009m), copyright 2009.)

Interestingly, *Fermi* LAT observations have also revealed γ-ray emission from four radio-loud narrow-line Seyfert 1 (NLS1) galaxies (Abdo *et al* 2009n, 2009o): PKS 1502+036 (z = 0.409), 1H 0323+342 (z = 0.061), PKS 2004-447 (z = 0.24) and PMN J0948+0022 (z = 0.585). These are the first detections of γ-ray emission from this class of sources. Unlike blazars that have strong relativistic jets and are hosted in elliptical galaxies, NLS1s are generally hosted in spiral galaxies. It is thought that the dichotomy of blazars (BL Lacs and FSRQs) and radio galaxies with strong, highly collimated relativistic jets on the one hand and Seyfert galaxies with relatively slow, weak and poorly collimated outflows on the other hand reflects galactic environmental differences such as the rate of accretion of gas or spin of the black hole. (e.g. Marscher 2009). The detection of high-energy radiation from NLS1s by *Fermi* is contrary to this paradigm. Abdo *et al* (2009o) have compared the jet powers of these sources with the jet powers seen in blazars and found them to be average. In addition these sources have small masses and high accretion rates (relative to the Eddington rate) compared to those inferred for blazars.

*4.2.2 Radio Galaxies*: Given that the γ-ray emission observed from blazars is likely produced in compact emission regions moving with relativistic bulk velocities in or near the parsec scale core in order to explain the observed rapid variability and to avoid attenuation due to pair production, it is natural to extend this picture to radio galaxies (Chiaberge *et al* 2001) that are believed to have jets oriented at systematically larger angles to the line of sight, thus constituting the parent population of blazars.

The nearest large radio galaxy, Cen A, may have been seen by EGRET. The Third EGRET catalog has a source positionally consistent with Cen A, and the energy spectrum appears to be a continuation of the spectrum seen at lower energies (Sreekumar *et al* 1999). In the absence of any variability correlated with other wavelengths, however, this identification was not certain. A similar situation existed for two other radio galaxies, NGC6251 (Mukherjee *et al* 2002) and 3C111 (Sguera *et al* 2005).

*Fermi* LAT observations have confirmed the discovery of Cen A (Abdo *et al* 2009p) and have detected Per A/NGC1275 (Abdo *et al* 2009q) and M87 (Abdo *et al* 2009r). Cen A has also been detected at TeV energies by H.E.S.S. (Aharonian et al 2009a). Unlike the dayscale variability seen at TeV energies (Acciari et al 2009a), there is so far no evidence for variability in the *Fermi* observations of the MeV/GeV emission in M87.

M87 is the faintest γ-ray radio galaxy detected so far with a >100 MeV flux (∼ 2.5 x $10^{-8}$ ph $cm^{-2}$ $s^{-1}$) about an order of magnitude lower than in Cen A and Per A; the corresponding > 100 MeV luminosity, 4.9 x $10^{41}$ erg $s^{-1}$, is 4 times greater than that of Cen A, but more than 200 times smaller than in Per A. The γ-ray photon index of M87 in the LAT band is similar to that of Per A ($\alpha$ = 2.3 and 2.2, respectively), while being smaller than observed in Cen A ($\alpha$ = 2.9).

Continued LAT monitoring of radio galaxies, coordinated with multi-wavelength observations, can extend the current study of 'quiescent' emission to possible flaring, in order to further address the physics of the radiation zone. While the extragalactic high-energy γ-ray sky is dominated by blazars, the LAT detections certainly indicate a population of γ-ray radio galaxies. Other examples, including the possible associations with EGRET detections like NGC 6251 and 3C 111 await confirmation with the LAT, and more radio galaxies are expected to be detected at lower fluxes. This holds great promise for systematic studies of relativistic jets with a range of viewing geometries in the high-energy γ-ray window opened up by the *Fermi* LAT.

*4.2.3 Large Magellanic Cloud*: Cosmic-ray interaction processes that operate in the Milky Way are expected to operate in other normal galaxies, although most of these are too far away to be detectable.

At a distance of 50 kpc, the Large Magellanic Cloud (LMC) is an exception, having been detected as an extended γ-ray source by EGRET (Sreekumar *et al* 1992). Based on 11 months of observations, *Fermi* LAT has detected the LMC at high significance with an integrated photon flux (> 100 MeV) of $(2.6 \pm 0.2) \times 10^{-7}$ ph cm$^{-2}$ s$^{-1}$ that corresponds to an energy flux of $(1.6 \pm 0.2) \times 10^{-10}$ erg cm$^{-2}$ s$^{-1}$ (Abdo *et al* 2010a). The emission maximum is located in the 30 Doradus star forming region and is consistent with the position of the Crab-like pulsars PSR J0540-6919 and PSR J0537-6910, and extended emission with a width of ~1.2°. However, the evidence for pulsed emission from either pulsar is marginal at best; there is a hint (2.4σ) of pulsations from PSR J0540-6919 at the expected pulse period. If γ-ray emission indeed originates from the pulsar, it implies a surprisingly large isotropic γ-ray luminosity conversion efficiency of $\eta_\gamma \approx 9\%$. In any case, the pulsar contribution to the γ-ray emission from 30 Doradus is minor so cosmic-ray interactions with the interstellar gas in the disk of the LMC and radiation fields are likely the dominant sources of emission. Abdo *et al* (2010a) find that the γ-ray emission correlates little with the gas density of the LMC and that the γ-ray morphology resembles more that of optical Hα line emission which traces regions of strong ionization. Assuming that the LMC disk emission can be attributed to cosmic-ray interactions with the gas, Abdo *et al* (2010a) have determined the ratio between the average cosmic-ray density in the LMC to that in the local interstellar medium to be $r_c = 0.22 \pm 0.08$. The γ-ray emission in the LMC shows little evidence for correlation with interstellar gas and apparently better traces regions of massive star formation, indicating that these regions are probable sites of cosmic-ray acceleration. The apparent tight confinement of the γ-ray emission to star forming regions in the LMC suggests a relatively short diffusion length for GeV protons.

*4.2.4 Starburst Galaxies*: If cosmic rays are accelerated by supernova remnant shocks that are formed when a star explodes, starburst galaxies, in particular, should have larger γ-ray intensities compared to the Milky Way due to the increased star-formation rate and greater gas mass, dust mass, and photon densities that serve as targets for γ-ray production by cosmic rays. Because cosmic rays diffuse throughout the Milky Way and make a bright γ-ray background glow, γ-rays are difficult to attribute to interactions of cosmic rays accelerated by Galactic supernovae. Direct evidence for the sources of cosmic rays is therefore still lacking. The supernova remnant origin for cosmic rays can also be tested, however, by measuring the γ-ray emission from star-forming galaxies. Indeed, estimates of the γ-ray luminosity of starburst galaxies made well before the launch of *Fermi* suggested that *Fermi* LAT should detect them (e.g., Ginzburg and Syrovatskii 1964; Hayakawa 1969; Akyüz, Broulliet and Ozel 1991; Völk, Aharonian and Breitschwerdt 1996; Romero and Torres 2003).

From the initial 11 months of sky-survey data, *Fermi* LAT has indeed detected the nearest starburst galaxies M82 and NGC 253, albeit NGC 253 with lower significance (Abdo *et al* 2010b). The reported flux levels (>100 MeV) are (1.6 ± 0.5 stat ± 0.3 sys) x $10^{-8}$ ph $cm^{-2}$ $s^{-1}$ for M82, and (0.6 ± 0.4 stat ± 0.4sys) x $10^{-8}$ ph $cm^{-2}$ $s^{-1}$ for NGC 253. The observed spectra are consistent with a power-law fit of index 2.2 ± 0.2 stat ± 0.05 sys for M82 and 2.0 ± 0.4 stat ± 0.05 sys for NGC 253, respectively (Abdo *et al* 2010b). Neither source showed any evidence of variability. Recently, the TeV detections of NGC 253 (Acero *et al* 2009) and M82 (Acciari *et al* 2009b) by H.E.S.S. and VERITAS, respectively, have been reported.

Relatively high star formation rates are observed within the central several hundred parsecs of these galaxies, where significant amounts of molecular gas and dust are also found. Estimates of the SN rate vary from ~ 0.08–0.3 SN $yr^{-1}$ in M82, to 0.1-0.3 SN $yr^{-1}$ in NGC 253, compared to the SN rate of ~ (1/50) SN $yr^{-1}$ in the Milky Way, or ~(1/200) SN $yr^{-1}$ in the LMC.

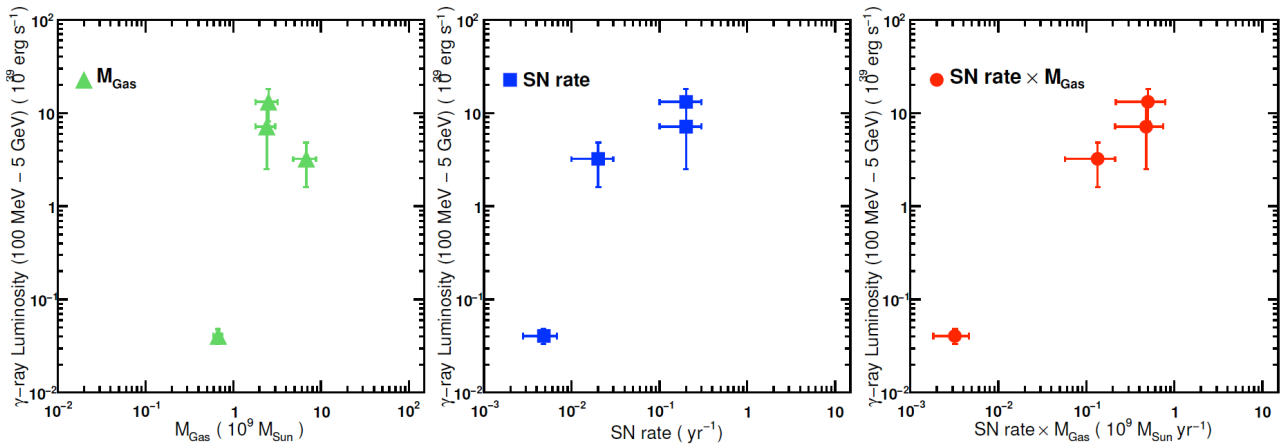

**Figure 13.** Relationship between supernova rate, total gas mass, and total γ-ray luminosity of four galaxies detected by their diffuse γ-ray emission (>100 MeV). The galaxies are, in order of ascending γ-ray luminosity, the LMC, Milky Way, NGC 253, and M82. Three panels shown compare different possible correlations with the γ-ray luminosity: total gas mass (left), supernova rate (center), and product of the total gas mass and supernova rate (right). (Reprinted by permission from the American Astronomical Society: *Astrophysical Journal* (Abdo *et al* 2010b), copyright 2009.)

Abdo *et al* (2010b) compare the SN rates, total galactic gas masses, and the γ-ray luminosities of these starburst galaxies with the LMC and the Milky Way. Figure 13 shows the product of SN rate and total galactic gas mass for each of these galaxies. The correlation spanning over two decades in both γ-ray luminosity and SN rate times gas mass is striking, indicating that supernova remnants are indeed the major sources of cosmic rays in normal galaxies. Still, as noted by Abdo *et al* (2010b), the exact details regarding cosmic-ray acceleration and propagation are unique to each individual galaxy. Radio and infrared observations show that the starburst activity in M82 and NGC 253 takes place in a relatively small central region and thus the distribution of the cosmic-ray particles in the galaxies is non-uniform as is the distribution of supernova explosions. In the case where γ-ray emission can be resolved, this situation can be seen directly, for example, in the LMC as discussed in the previous section. Using the total gas mass of each galaxy and assuming uniform supernova explosion rate throughout might provide a qualitative description of underlying mechanisms behind cosmic-ray enhancement in the interstellar medium. However, quantitative understanding requires detailed modeling and observational feedback from this and a larger sample of objects.

*4.2.5 Gamma-Ray Bursts*: During the past decade the study of X-ray, optical, and radio afterglows of GRBs has revealed their distance scale, helping to transform the subject from phenomenological speculation to quantitative astrophysical interpretation. We now know that long-duration GRBs and at least some short-duration GRBs lie at cosmological distances and that both classes involve extremely powerful, relativistic explosions.

The current picture that has emerged of GRB physics is that an initial fireball powers a collimated, super-relativistic blast wave with initial Lorentz factor $\sim 10^2$–$10^3$. Prompt *γ*-ray and X-ray emission from this "central engine" may continue for a few x $10^3$ s or longer. Then external shocks arising from interaction of the ejecta with the circumstellar environment at lower Lorentz factors give rise to afterglows in the X-ray and lower-energy bands that are detected for hours to months. The physical details—primary energy source and energy transport, degree of blast wave collimation, and emission mechanisms—continue to be debated (Zhang and Meszaros 2004).

Simulations, based on extrapolations from the BATSE-detected GRBs (Preece *et al* 2000), and adopting the distribution of Band parameters of the catalog of bright BATSE bursts (Kaneko *et al* 2006), suggested that the LAT should detect between one burst per week and one burst per month, depending on the GRB model for high-energy emission (Atwood *et al* 2008). The observed rate of LAT-detected bursts is about one per month. In the first 13 months of operations, *Fermi* LAT has detected high-energy emission from the 10 GRBs listed in table 2. The *Fermi* GBM also detected all of these bursts at lower energies and six of them triggered follow-up observations with the Swift satellite (Gehrels *et al* 2004) within 24 hours. The Swift localizations permitted optical follow-up observations; resulting in redshift determinations for 50% of the LAT detected bursts. Note that two of the LAT-detected bursts are short-duration bursts despite not so optimistic predictions of high-energy emission from this class of bursts (Nakar 2007).

There have been several notable results from the *Fermi* LAT detections of high-energy emission from GRBs:

(i) During the prompt phase of most of the bursts, the onset of the GeV emission is delayed relative to the low-energy (MeV) emission. The case of GRB 080916C, shown in figure 14, is illustrative (Abdo *et al* 2009s).

(ii) The high-energy emission from almost all LAT-detected bursts (except GRB 090217) persists longer than the emission in the keV-MeV band. For example, GRB 080916C showed significant high-energy emission up to 1,400 s after the trigger. For GRB 090510, the first burst ever observed simultaneously from the optical to GeV γ-rays (De Pasquale *et al* 2010), the high-energy emission persisted for ~200 seconds.

| burst | characteristics |
|---|---|
| GRB 080825C | long duration; weak |
| GRB 080916C | long duration; intense; z = 4.35 |
| GRB 081024B | short duration; weak |
| GRB 081215A | long duration; 86° to the LAT boresight |
| GRB 090217 | long duration; featureless light curve |
| GRB 090323 | long duration; ARR; z = 3.6 |
| GRB 090328 | long duration; ARR; z = 0.74 |
| GRB 090510 | short duration; ARR; z = 0.903 |
| GRB 090626 | long duration |
| GRB 090902B | long duration; intense; ARR; z = 1.82 |

**Table 2.** Gamma-ray bursts detected by the *Fermi* LAT between August 2008 and September 2009. Each of these bursts was also detected by the *Fermi* GBM. Redshifts have been obtained for half of these bursts. Four of the bursts caused an autonomous repoint request (ARR) to be generated and initiated. All of the bursts, except GRB 090217, exhibited temporally extended high-energy emission that lasted from tens of seconds to thousands of seconds

(iii) High-energy emission observed on short timescales in most of the LAT-detected bursts, combined with the requirement that the opacity for high-energy photons be sufficiently low that the radiation can escape from the source and be observed, leads to the highest lower limits, typically $\Gamma > 1,000$, on the Lorentz factor of the bulk outward flow of the emitting material. For example, in the case of the short burst GRB 090510, a 33 GeV photon emitted during the first second of the burst leads to the highest lower limit of $\Gamma > 1,200$ (Abdo *et al* 2009t), suggesting that the outflows powering short GRBs are at least as highly relativistic as those powering long GRBs. It is not yet clear if high-energy emission is always accompanied by large Lorentz factors.

(iv) GRB 080916c, at redshift z = 4.35, also had an apparent isotropic energy release of $E_{iso}$ ~ 8.8 x $10^{54}$ ergs, that is nearly 5 times the Solar rest energy and the highest inferred for any GRB to date (Abdo *et al* 2009s). On energetic grounds, this suggests that the GRB outflow powering this emission was collimated into a narrow jet.

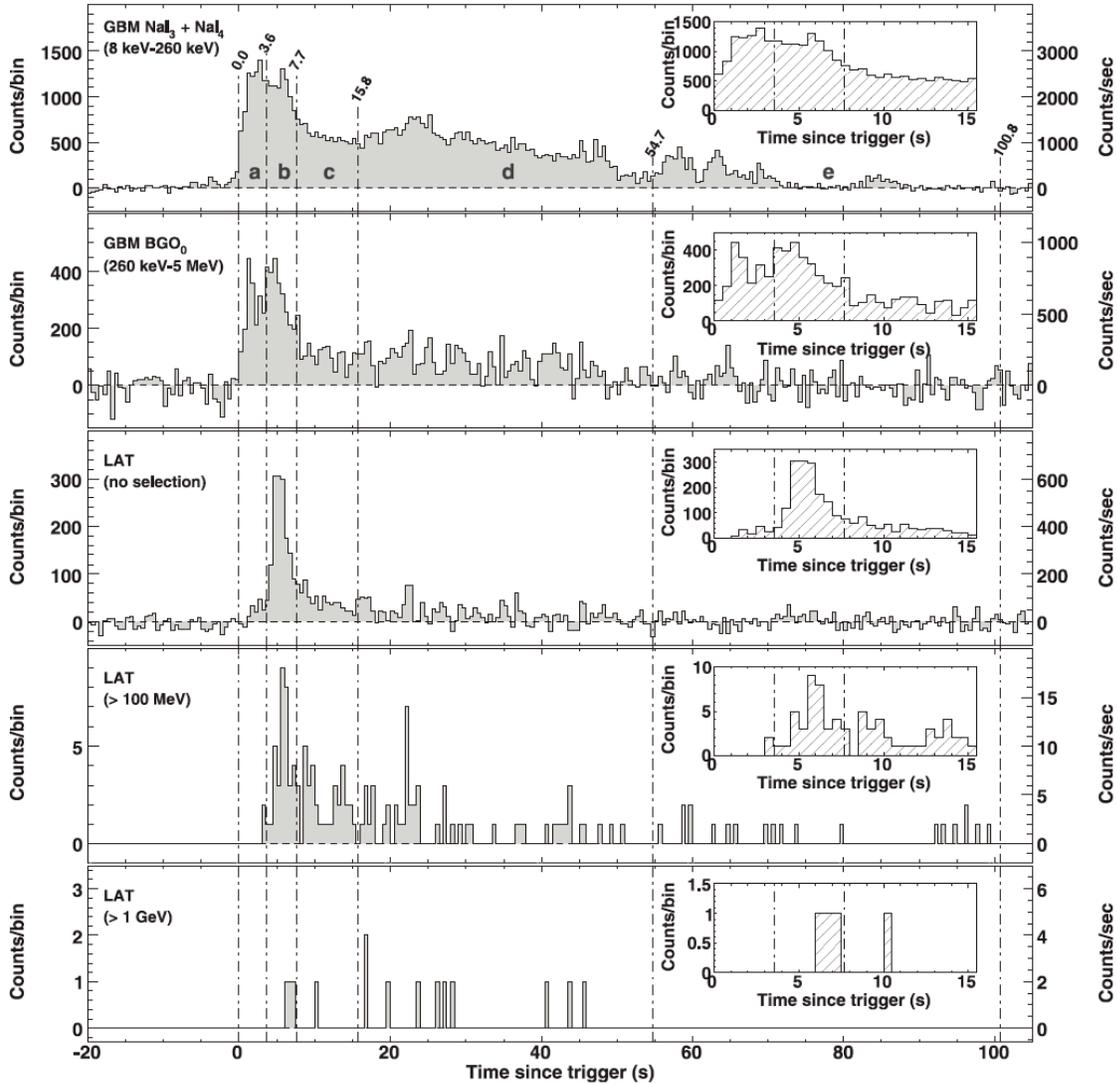

**Figure 14.** Light curves for GRB 080916C observed with the GBM and the LAT, from lowest to highest energies. The top three graphs represent the background subtracted light curves for the GBM NaI, the GBM BGO, and the LAT. The third shows all LAT events passing the onboard event filter for γ-rays that have at least a reconstructed track found in the ground analysis. (Insets) Views of the first 15 s from the trigger time. In all cases, the bin width is 0.5 s; the per-second counting rate is reported on the right for convenience. (Reprinted by permission from the American Association for the Advancement of Science: *Science* (Abdo *et al* 2009s), copyright 2009.)

(v) The large range of photon energies, large distances, and timescales of some of the bursts allow an experimental check of the assumption that all photons travel at the same speed in vacuum. Some quantum gravity models suggest that velocity dispersion may indeed exist (Amelino-Camelia *et al* 1998). If the dispersion is linear, the difference in the arrival times $\Delta t$ may be characterized as the ratio of photon energy difference to the quantum gravity mass scale, $\Delta E/M_{QG}$, and of course depends on the distance the photons traveled. Smaller time differences imply larger values of $M_{QG}$, and the interesting scale is set by the Planck mass, $1.22 \times 10^{19}$ GeV/c$^2$. For GRB 080916C, the arrival time of a 13 GeV photon 16.54 seconds after the burst trigger provides a conservative upper limit on its $\Delta t$ relative to ~ MeV photons and therefore a *lower* limit on the quantum gravity mass, $M_{QG} > 1.3 \times 10^{18}$ GeV/c$^2$ (Abdo *et al* 2009s), only one order of magnitude smaller than the Planck mass, $M_{Pl} = 1.22 \times 10^{19}$ GeV/c$^2$. The main assumption for the lower limit is that the high-energy emission at the source did not occur *earlier* than the low-energy emission. The most recent constraint provided by the short burst GRB 090510, at z=0.9 with a ~31 GeV photon detected at 0.829 s, is the most stringent to date with a limit $M_{QG} > \sim 1.2\ M_{Pl}$ again assuming a linear dispersion relation (Abdo *et al* 2009t).

*4.2.6 Diffuse isotropic radiation*: While the diffuse Galactic emission is produced by interactions of cosmic rays, mainly protons and electrons, with the interstellar gas and radiation field, the much fainter extragalactic background (EGB) is the sum of contributions from unresolved sources and truly diffuse emission, including possible signatures of large scale structure formation, annihilation of cosmological dark matter, emission produced by ultra-high-energy cosmic-rays interacting with relic photons, and many other processes.

The Fermi LAT Collaboration has reported a measurement of the spectrum of the isotropic diffuse γ-ray radiation from 200 MeV to 100 GeV (Abdo *et al* 2010c). The biggest challenge in the determination of the EGB is the subtraction of the various strong foregrounds that exist in the LAT photon dataset. This is also the source of the largest systematic uncertainty. Most important are the contributions from the Galactic diffuse emission, the background from misclassified cosmic rays and from the resolved sources. The instrumental background is suppressed by applying very stringent event selection criteria beyond the standard event selection. The diffuse emission originating from cosmic-ray interactions with the interstellar medium in the Milky Way and the contribution from point sources is fitted to the Fermi LAT dataset.

The spectrum found by Abdo *et al* (2010c), shown in figure 15, is a featureless power law, significantly softer than the one obtained from EGRET observations by Sreekumar *et al* (1998). A possible reason for the discrepancy with the EGRET measurement might be an overestimation of the flux in the EGRET analysis for energies > 1 GeV as indicated by the difference between EGRET and LAT spectra for the intermediate latitude region (Abdo *et al* 2009b) as well as the spectrum of the Vela pulsar (Abdo *et al* 2009e). Also, the spectrum does not show a distinctive peak at E >3 GeV found in a reanalysis of the EGRET data with an updated diffuse model (Strong, Moskalenko and Reimer 2004), which had been attributed to a possible contribution of dark matter (Elsässer and Mannheim 2005).

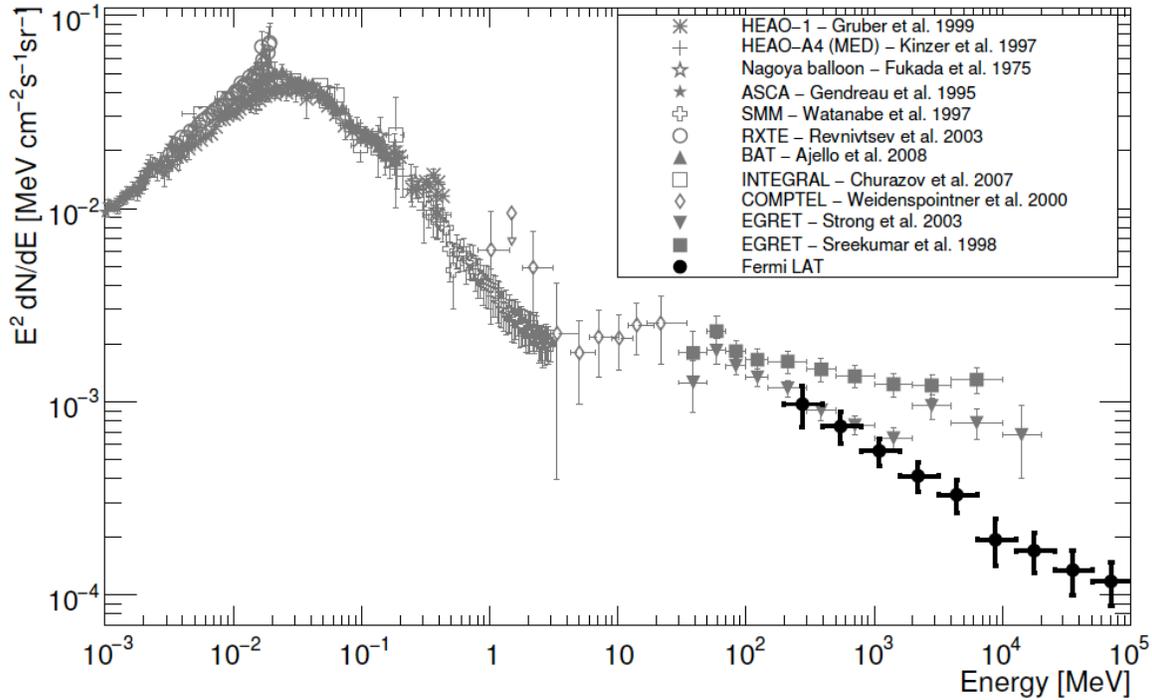

**Figure 15.** Spectral energy distribution of the extragalactic diffuse emission between 1 keV and 100 GeV measured by various instruments, including the *Fermi* LAT. For references to the various observations see Ajello *et al* (2008). Adapted from Abdo *et al* (2010c)

*4.3 Local Sources; the Sun and the Moon*: EGRET observed a long-duration high-energy solar event on 1991 June 11 when >50 MeV emission was detected for a duration of over 8 hr (Kanbach *et al* 1993). Pion-decay γ-rays appeared to dominate the emission.

To date, *Fermi* has not detected such solar events. Before launch, solar activity was expected to rise in 2008 with a peak occurring as early as 2011. However, solar activity has been anomalously low. Despite the non-detections of high-energy solar flare events, *Fermi* has detected γ–ray emission from the quiescent Sun (Orlando 2009) due to cosmic-ray proton interactions with the solar atmosphere at a level consistent with estimates made by Seckel *et al* (1991) and extended heliospheric emission due to inverse-Compton interactions of cosmic-ray electrons and positrons with the solar radiation field (Moskalenko, Porter and Digel 2006, 2007; Orlando and Strong 2008).

The Moon is also a source of γ-rays due to cosmic-ray interactions with its surface and has been detected by EGRET (Thompson *et al* 1997) and is seen by the *Fermi* LAT as well (Giglietto 2009). Unlike the cosmic-ray interactions with the gaseous atmospheres of the Earth and the Sun, the Moon surface is solid, consisting of rock, leading to a spectrum of γ-rays that is very steep with an effective cutoff around 3–4 GeV, 600 MeV for the inner part of the Moon disk (Moskalenko and Porter 2007).

*4.4 Dark Matter searches:* Strong evidence for the presence of large amounts (3 to 4 times that of ordinary matter) of non-baryonic matter in the universe is provided by its gravitational interaction with ordinary matter; *i.e.*, the rotation curves of galaxies, structure-formation arguments, the dynamics and weak lensing of clusters of galaxies, and *WMAP* measurements of the cosmic microwave background (Spergel *et al* 2007; for review, see, *e.g.*, Bergström 2000). In the scenario where dark matter is of an elementary particle nature, annihilations and/or decays of these as yet undetected particles could produce an observable signal. The detection of such a signal is dubbed an "indirect" detection in that one is seeing the remnants of the final state resulting from annihilations or decays. However, estimating the magnitude along with the other characteristics of such a signal requires many assumptions and hence these estimates vary by orders of magnitude. The search for such signals is an important objective of the *Fermi* mission.

The γ-ray flux from dark matter annihilations can be written as (Bergstrom, Ullio and Buckley 1998)

$$\phi(E,\psi) = \frac{\langle\sigma v\rangle}{4\pi} \sum_f \frac{dN_f}{dE} B_f \int_{l.o.s.} dl(\psi) \frac{1}{2}\left(\frac{\rho(l)}{M_\chi}\right)^2$$

where $\phi$ is the observed flux at energy $E$ observed in direction $\psi$. $<\sigma v>$ is the annihilation cross-section times the velocity. The sum is over all final states with photons where $B_f$ are the branching fractions and $dN_f/dE$ are the spectra for each final state. The integral sums the number density squared along the line-of-sight. Each factor in this expression is uncertain at the *order-of-magnitude* (or several) level! In the case of dark matter decays, simply substitute one over the decay time ($1/\tau$) for $<\sigma v>$ and eliminate the square on the dark matter density factor.

γ-ray signals originating from dark matter are being looked for in the first year *Fermi* LAT data using many different approaches. These include line searches and studies of emission from the Galactic Center, halo objects, and galaxy clusters, along with diffuse galactic and extragalactic γ-ray radiation.

*4.4.1 Dark Matter Line Searches:* In the case of two dark matter particles annihilating into a two body final state with one of the particles a photon, the photon will have a well-defined energy (Rudaz and Stecker 1991). Two such possibilities have been identified: γ–γ final states and γ-$Z^o$ boson final states. Estimates of the branching ratio for these states are highly uncertain and vary by several orders of magnitude. In many models the branching ratio for annihilation into lines is very small ($\sim 10^{-3}$ to $10^{-4}$) and would be essentially undetectable with the LAT (Baltz *et al* 2008) although some models predict much stronger lines (e.g., Gustafson *et al* 2007). Nevertheless a "line" like feature at high energy would be the clearest and most definitive signature of dark matter particle annihilations and given the large uncertainties it is attractive to search for.

The *Fermi* LAT Collaboration has used two regions of the sky to search for lines (Abdo *et al* 2010d): region A is an all-sky region with the Galactic plane removed (*i.e.*, $|b| > 10^o$) and region B includes region A plus a $20^O$ x $20^O$ square region centered on the Galactic center. These regions were chosen in hopes of improving the signal-to-noise by eliminating the very bright, high-energy galactic plane emission. The Galactic Center is included in region B since dark matter is thought to be most concentrated at this location. All *Fermi* LAT point sources within these regions were eliminated by excluding a circular region (radius = $0.2^o$) around each source. Note that the LAT point-spread function at 20 GeV is $\sim 0.1^o$ and this is the lower limit for the energy band used in these searches. The upper limit of the energy band was 300 GeV due to limited statistics above this energy.

No significant line feature has been found. The 95% confidence upper limit on the presence of a line like feature for energies $\geq 60$ GeV is found to be less than $\sim 2 \times 10^{-9}$ cm$^{-2}$ s$^{-1}$ for both regions.

*4.4.2 Searches for dark matter in Galactic halo objects*: In the halo of the Galaxy two possible sources for enhanced dark matter annihilation are being examined: (i) clumps as predicted by N-Body simulations and (ii) dwarf spheroidal galaxies (dSph). Halo clumps of dark matter predicted by N-Body simulations are certainly more speculative than the known locations of high concentrations of dark matter in several dSph, inferred from studies of stellar motion, that are in close proximity to the Milky Way.

Large N-body simulations such as Via-Lactea II (Kuhlen *et al* 2008) and Aquarius 5 (Springel *et al* 2008) predict that DM will tend to clump as opposed to remain uniformly distributed within a galaxy. They predict a distribution in masses for these clumps with the frequency of clumping increasing as the size of the clump becomes smaller. Overall they predict that the total rate for DM annihilations will be *boosted* by perhaps a factor of 4 to as much as 15 over what would result from a uniform distribution. Also, if there were to be large and somewhat close-by clumps, they could be visible in γ-rays from the ongoing annihilations. The sensitivity and hence the rate at which these clumps might be found is very model dependent, but overall one to two might be seen with a one year exposure of the sky.

These dark matter clumps would appear as unassociated γ-ray sources, with a spectrum characteristic of the annihilation process. If close by, they would also appear to be spatially extended. Searches are underway in the LAT data for such occurrences. The search criteria are 1) no counter-part observed close to the candidate location, 2) the emission is constant in time, 3) they be spatially extended ($\sim 1^o$), and 4) their spectrum is consistent with expectations for γ-rays from DM. In the data from the first 3 month, one candidate did emerge which approximately satisfied these four criteria. Of course all such early detections are best tested by an increase in significance as more data becomes available. After 10 months of data, the proposed DM clump has started to become resolved into two nearby point sources. Hence the LAT claims no significant detection to date. However the analysis is ongoing.

Dwarf Spheriodal Galaxies (DSphs) are an excellent target for dark matter searches. These clusters of stars self gravitate and orbit about our Galaxy. The motions of the stars contained in these objects can be measured and used to determine the total gravitating mass. The total luminous mass is determined by summing the total stellar mass detected optically. In the most promising candidates for dark matter detection the ratio of total mass to luminous mass exceeds $\sim 10^3$ albeit often with large errors. In many cases this ratio exceeds $10^2$.

To date, the LAT search for a dark matter signal in DSphs has focused on a selection of 10 DSphs. These DSphs are all located within 150 kpc of the Sun and are more than $30^o$ off of the Galactic plane. These criteria were imposed to maximize any signal as well as reduce the foreground γ-ray background from the Milky Way. From the first 9 months of data, no significant detections of any of these 10 objects were made even before imposing a requirement that the spectral content "look like" dark matter. The 95% flux limits integrated above 100 MeV are all at the ~few x $10^{-9}$ ph cm$^{-2}$s$^{-1}$ level. These limits will slowly improve with time.

*4.4.3 Search for dark matter in galaxy clusters:* Analogous to the DSph search, another "calibrated" dark matter source is galaxy clusters. These are bunches of rather tightly grouped galaxies. One then looks at the their relative motion with respect to each other to infer the amounts of missing matter not explained by their luminosities.

Two such clusters have been examined using the first 9 months of Fermi LAT data; the Fornax Cluster and the Coma Cluster. No γ-ray signal has been seen from either.

*4.5 Cosmic-ray electrons and positrons*: Although designed as a high-sensitivity γ-ray telescope, the *Fermi* LAT is also an excellent electron detector with a very large acceptance, exceeding 2 m$^2$ sr at 300 GeV. Building on the γ-ray analysis, the LAT Collaboration has developed an efficient electron detection strategy that provides sufficient background rejection for measurement of the steeply falling electron spectrum up to 1 TeV (Abdo *et al* 2009u). Of course *Fermi* cannot directly distinguish electrons and positrons. The *Fermi* LAT results in figure 16 show that the electron-positron spectrum falls with energy as ~$E^{-3.04}$ and does not exhibit prominent spectral features. Prior to 2008, the high-energy electron spectrum was measured by balloon-borne experiments (Nishimura *et al* 1980; Kobyashi *et al* 2004) and by a single space mission, AMS-01 (Aguilar *et al* 2002). The measured fluxes differed by factors of 2 to 3.

The conventional diffusive cosmic-ray model (Moskalenko and Strong 2001), dashed curve in figure 16, is based on the assumption that electrons originate from a distribution of distant sources mainly associated with supernova remnants and pulsars. Positrons are produced as secondaries resulting from the interaction of protons with the interstellar medium and are predicted to be a small fraction (~5%) of the total flux of electrons and positrons. The conventional model predicts a featureless spectrum from 10 GeV up to few hundreds of GeV. Above that energy, due to the actual stochastic nature of electron sources in space and time, and to the increasing synchrotron and inverse Compton energy losses, the spectral shape may exhibit spatial variations on a scale of a few hundred parsecs. Also nearby sources will start contributing significantly to the observed local flux and may induce important deviations from a simple power law spectrum (Nishimura *et al* 1980; Kobayashi *et al* 2004; Aharonian *et al* 1995; Pohl and Esposito 1998).

The ATIC (Chang *et al* 2008) and PPB-BETS (Torii *et al* 2001) balloon experiments have reported detections of a prominent spectral feature at ~500 GeV in the total electron plus positron spectrum (see figure 16). A prominent feature in the spectrum is not seen in either the *Fermi* LAT data or in the results reported by the H.E.S.S. collaboration (Aharonian *et al* 2008b, 2009b). The H.E.S.S. spectrum steepens significantly above 600 GeV. While the *Fermi* data in figure 16 suggests a deviation from a flat spectrum, it is consistent with a power-law if systematic errors are conservatively added point-to-point in quadrature with statistical errors (Abdo *et al* 2009u).

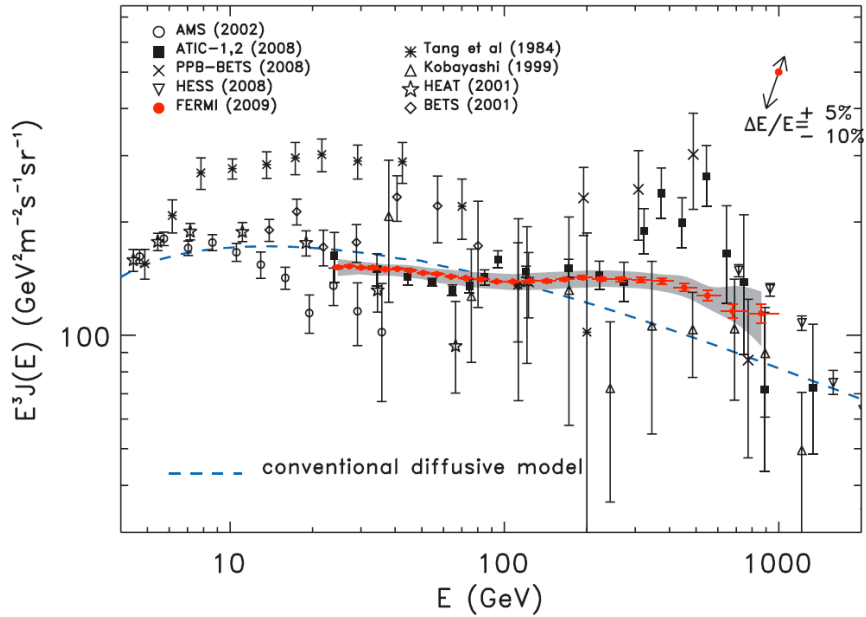

**Figure 16.** The *Fermi* LAT cosmic-ray electron-positron spectrum (red filled circles). The gray band shows systematic errors. The two-headed arrow in the top-right corner of the figure gives size and direction of the rigid shift of the spectrum implied by a shift of +5%, -10% of the absolute energy, corresponding to the present estimate of the systematic uncertainty of the LAT energy scale. Other high-energy measurements and a conventional diffusive model (Strong, Moskalenko and Reimer 2004) are shown. (Reprinted by permission from the American Physical Society: *Physical Review Letters* (Abdo *et al* 2009u), copyright 2009.)

The Pamela satellite experiment, with a magnetic spectrometer, has reported measurements of the spectrum of the positron fraction, $e^+/(e^+ + e^-)$, that shows an excess above ~10 GeV that increases with increasing energy up to at least 100 GeV (Adriani *et al* 2009), and is not explained by secondary production if the electron spectrum is as hard as the *Fermi* measurements suggest.

Abdo *et al* (2009u) point out the *Fermi* LAT observation that the electron spectrum is much harder than the conventional one may be explained by assuming a harder electron spectrum at the source, which is not excluded by other measurements. They also suggest that the significant flattening of the LAT data above the model predictions for $E \geq 70$ GeV could indicate the presence of one or more local sources of high-energy cosmic-ray electrons.

The Pamela results along with the *Fermi* and H.E.S.S. results may indicate the presence of a nearby primary source(s) of electrons and positrons, two classes of which stand out: nearby pulsar(s) (e.g. Shen 1970; Aharonian *et al* 1995; Kobayashi *et al* 2004; Yuksel, Kistler and Stanev 2009; Grasso *et al* 2009) and dark matter, either by annihilation (e.g., Zhang and Cheng 2001; Malyshev, Cholis and Gelfand 2009; Cholis, Goodenough and Weiner 2009; Bergström, Bringmann and Edsjö 2008; Arkani-Hamed *et al* 2009; Cirelli *et al* 2009) or decay (Yin *et al* 2009; Hamaguchi *et al* 2009), for example through grand unified interactions with a lifetime of order ~$10^{26}$ s (e.g. Arvanitaki *et al* 2009).

# 5. Summary: What next from *Fermi*

In the conclusion of a review of EGRET results, Thompson (2008) presented a list of open questions left behind by EGRET. Table 3 summarizes these questions and the progress *Fermi* has made in answering them. Starting with the second year of operations, all *Fermi* photon data and a set of analysis tools are public immediately after low-level data processing[3]. With a 5-year nominal science mission and a goal of 10 years of operation, we look forward to many new results from *Fermi* in the future.

| Question | Status |
|---|---|
| What is the nature of the diffuse Galactic γ-ray radiation, and in particular the GeV excess? | *Fermi* does not find GeV excess at mid-galactic latitudes. Understanding of origin of diffuse radiation still incomplete; spatial mapping in much finer detail and room for fundamental discoveries above 100 GeV. |
| Does the γ-ray radiation from the Milky Way or its surroundings contain clues to unseen forms of matter, such as cold dark gas or dark matter? | Underway; requires significantly more exposure, particularly above 100 GeV, and deeper understanding of conventional astrophysical sources, to probe the most interesting discovery region for dark matter. |
| What will a larger sample of γ-ray pulsars reveal about the location of the particle acceleration and the particle interaction processes under extreme conditions? | *Fermi* observations of the spectrum of pulsars, particularly Vela, favor outer-magnetosphere emission models. Detections of many pulsars with precise γ-ray lightcurves vs energy will test three-dimensional magnetic field models and map emission mechanisms. |
| How many radio-quiet pulsars will be found, and what will those pulsars say about the neutron star population of our Galaxy? | A large fraction (~37%) of the *Fermi*-detected pulsars, not including millisecond pulsars, are "radio-quiet". Sample large enough to constrain contribution from unresolved pulsars to diffuse emission. |
| Which binary systems produce γ-rays, and how do those systems work? | Three binaries, LS I +61°303, LS 5039 and Cygnus X-3, detected by *Fermi* (as well as by TeV telescopes). Emission likely arises from electrons accelerated near a neutron star or a black hole. |
| What other classes of galactic and extragalactic objects have enough energy to produce γ–rays detectable by the new generation (e.g. *Fermi*) of telescopes? | New galactic GeV source classes: globular clusters, HMXBs, SNRs and Pulsar Wind Nebula, evidence of Galactic transients; new extragalactic source classes: starburst galaxies (e.g. M82, NGC 253), definite detection of radio galaxies (Cen A, M87, Per A) |
| Will including new, high-quality γ-ray measurements of blazar spectra and time variability in multiwavelength studies provide the clues to jet properties such as composition and possibly to jet formation, collimation, | Still open. High-quality monitoring of γ-ray emission from large number of blazars has been done since the start of operations, and it will continue. |

---

[3] data can be obtained from the Fermi Science Support Center at
http://fermi.gsfc.nasa.gov/ssc/

| and γ-ray emission region? | Multiwavelength studies are ongoing. |
|---|---|
| What will the new γ-ray measurements of other galaxies tell us about cosmic rays and matter densities in these systems? | Observations of γ-ray emission from starburst galaxies and LMC open the door to more detailed understanding of cosmic ray production. |
| Will the new data resolve the diffuse extragalactic radiation as a collection of discrete sources, or will there be some residual diffuse emission that demands a new and possibly exotic explanation? | Integrated unresolved isotropic flux ~70% less than that measured by EGRET, consistent with number-flux distribution observed for blazar sources. |
| Do most or all GRBs have high-energy emission, and what does that radiation say about the forces at work in these explosive phenomena? | Of the 138 bursts in LAT FoV during the 1st year, 10 had detectable high-energy emission, including 2 short GRBs. LAT detections indicate extreme relativistic expansion, delayed high-energy emission |
| Can high-energy γ-ray measurements of solar flares shed new light on solar activity? | No high-energy flares yet detected by *Fermi*. Now entering the more active time in the 11-year solar cycle. |
| What surprises will be found in the γ-ray sky? | The sky is full of γ-ray-only pulsars; the cosmic ray $e^+ + e^-$ energy distribution is different from expectation, requiring further investigation; there is high-energy emission from short GRBs; stay tuned for more from *Fermi* |

**Table 3.** Summary of scientific legacy questions left by EGRET and the progress made in answering them with *Fermi*.

## 6. Acknowledgments

The first-year results summarized here were made possible by the extraordinary skill and dedication of the entire Fermi Gamma-ray Space Telescope team and the sustained support of the sponsoring agencies in the United States, France, Germany, Italy, Japan and Sweden. In addition there there has been significant participation from observatories in Great Britain and Australia.

The *Fermi* LAT Collaboration is supported by NASA and the Department of Energy in the United States; the Commissariat à l'Energie Atomique and CNRS/Institut National de Physique Nucléaire et de Physique des Particules in France; the Agenzia Spaziale Italiana, Istituto Nazionale di Fisica Nucleare, and Istituto Nazionale di Astrofisica in Italy; the Ministry of Education, Culture, Sports, Science and Technology, High Energy Accelerator Research Organization (KEK), and Japan Aerospace Exploration Agency in Japan; and the K. A. Wallenberg Foundation, Swedish Research Council, and National Space Board in Sweden.

The *Fermi* GBM Collaboration is supported by NASA in the United States and Deutsches Zentrum für Luft-und Raumfahrt in Germany.

In particular, the authors thank E. Bloom, S. Digel, F. Longo, J. McEnery, I. Moskalenko, N. Omodei, F. Piron, R. Rando, D. Smith, A. Strong, and D. Thompson for useful comments about this manuscript.